\documentclass[a4paper, 5p, sort&compress]{elsarticle}%5p for double column
\usepackage{ucs}
\usepackage[utf8]{inputenc}
\usepackage{amsmath,amssymb,amsfonts,mathtools}
\usepackage[english]{babel}
\usepackage{fontenc}
\usepackage{graphicx}
\usepackage{ulem}
\usepackage[margin=0pt,font=footnotesize, labelsep=colon]{caption}
\usepackage{tabularx}
\usepackage{multirow}
\usepackage{units}
\usepackage{booktabs}
\usepackage{bbm}
\usepackage{eurosym}
\usepackage{color}
\usepackage[table]{xcolor}
\usepackage{colortbl}
\usepackage{supertabular}
\usepackage{array,lipsum}
\definecolor{rgrey}{gray}{0.75}
\newcommand{\mean}[1]{\langle #1 \rangle}

\newcommand{\fref}[1]{Figure~\ref{#1}}

\newcommand{\highlight}[1]{{\color{black}#1}}

\usepackage[noabbrev]{cleveref}
\usepackage[export]{adjustbox}[2011/08/13]
\newcommand{\paren}[1]{\left(#1\right)}
\newcommand{\chromowidth}{1.00 \columnwidth}
\graphicspath{{Figures/}}
\usepackage{algorithm}
\usepackage[noend]{algpseudocode}
\usepackage{placeins}
\usepackage{subcaption}
\usepackage[toc,nonumberlist]{glossaries}
\usepackage{float}
\newfloat{algorithm}{H}{lop}
\floatname{algorithm}{Algorithm}

\newacronym{vres}{VRES}{Variable Renewable Energy Sources}
\newacronym{rea}{REA}{Renewable Energy Atlas}
\newacronym{ncep}{NCEP}{American National Centers for Environmental Prediction}
\newacronym{cfsr}{CFSR}{Climate Forecast System Reanalysis}
\newacronym{ac}{AC}{Alternating Current}
\newacronym{dc}{DC}{Direct Current}
\newacronym{lcoe}{LCOE}{Levelised Cost of Electricity}
\newacronym{hvdc}{HVDC}{High Voltage Direct Current}
\newacronym{iset}{IWES}{Fraunhofer-Institut f\"ur Windenergie und Energiesystemtechnik}
\newacronym{ga}{GA}{Genetic Algorithm}
\newacronym{gas}{GAS}{Greedy Axial Search}
\newacronym{cs}{CS}{Cuckoo Search}
\newacronym{de}{DE}{Differential Evolution}
\newacronym{CapEx}{CapEx}{capital expenditures}
\newacronym{OpEx}{OpEx}{operational expenditures}

\begin{document}
\crefname{appendix}{}{}

%*************************************************
\begin{frontmatter}

\title{
Optimal heterogeneity in a simplified highly renewable European electricity system
}

\author[label1]{Emil H. Eriksen}
\ead{emher@au.dk}
\author[label1]{Leon J. Schwenk-Nebbe}
\ead{leon@schwenk-nebbe.com}
\author[label2,label3]{Bo Tranberg}
\ead{bo@eng.au.dk}
\author[label4]{Tom Brown}
\ead{brown@fias.uni-frankfurt.de}
\author[label2]{Martin Greiner}
\ead{greiner@eng.au.dk}
\address[label1]{Department of Physics and Astronomy, Aarhus University, Ny Munkegade 120, 8000 Aarhus C,  Denmark}
\address[label2]{Department of Engineering, Aarhus University, Inge Lehmanns Gade 10, 8000 Aarhus C,  Denmark}
\address[label3]{Danske Commodities A/S, Vaerkmestergade 3, 8000 Aarhus C, Denmark}
\address[label4]{Frankfurt Institute for Advanced Studies (FIAS), Johann Wolfgang Goethe Universit\"at, Ruth-Moufang-Stra{\ss}e 1, 60438 Frankfurt am Main, Germany}

\begin{abstract}
The resource quality and the temporal generation pattern of variable renewable energy sources vary significantly across Europe. In this paper spatial distributions of renewable assets are explored which exploit this heterogeneity to lower the total system costs for a high level of renewable electricity in Europe. Several intuitive heuristic algorithms, optimal portfolio theory and a local search algorithm are used to find optimal distributions of renewable generation capacities that minimise the total costs of backup, transmission and renewable capacity simultaneously. Using current cost projections, an optimal heterogeneous distribution favours onshore wind, particularly in countries bordering the North Sea, which results in average electricity costs that are up to 11\% lower than for a homogeneous reference distribution of renewables proportional to each country's mean load. The reduction becomes even larger, namely 18\%, once the transmission capacities are put to zero in the homogeneous reference distribution. Heuristic algorithms to distribute renewable capacity based on each country's wind and solar capacity factors are shown to provide a satisfactory approximation to fully optimised renewable distributions, while maintaining the benefits of transparency and comprehensibility. The sensitivities of the results to changing costs of solar generation and gas supply as well as to the possible cross-sectoral usage of unavoidable curtailment energy are also examined.
\end{abstract}


\end{frontmatter}

%*************************************************************************
\section{Introduction}
\label{sec:one}
%*************************************************************************

The ambitious renewable energy targets set by European governments \cite{eu2050} imply that the share of renewables in electricity generation will increase significantly in the years to come.  At present, the leading renewable technologies are wind, solar photovoltaics (PV) and hydroelectricity, of which only wind and solar PV have the potential for large scale expansion. The uneven distribution of wind and solar resources across the continent raises the question of how best to exploit these heterogeneous resources. If wind and solar generation capacities are concentrated in those countries with the best resources, this may increase demand for transmission and increase energy imbalances between countries; if wind and solar generation are distributed homogeneously, then the best renewable resources will not be fully used and total system costs may be higher than the heterogeneous optimum. In this paper, the consequences of heterogeneity for the whole electricity system, including backup generation and transmission, will be quantified.

Since wind and solar PV are both \gls{vres}, backup generation is needed if the electrical demand is to be met at all times. Backup generation introduces additional system costs, which depend on the mismatch between \gls{vres} generation and load. Using the degrees of freedom associated with the choice of the capacity distributions of \gls{vres} for each country, it is possible to smooth out the aggregated temporal generation pattern or even shape it towards the load pattern. As a result, the mismatch and thus the backup requirements is lowered.  To decrease the dimensionality of the problem, renewable assets can be assigned homogeneously, proportional to the mean load of each country,  with a uniform wind-to-solar mixing factor. This approach is demonstrated in \cite{Heide2010,Heide2011}, where optimal wind-to-solar mixes for Europe are found that minimise balancing and storage costs. Further reductions in backup requirements are possible by extending the transmission network to enable more energy exchange between the countries \cite{Rodriguez2014,sarah}. The implications for total system costs of different homogeneous renewable penentrations, wind-solar mixes and transmission levels were considered in \cite{Sensitivity}, where the cost-optimal design was found to consist of a renewable energy penetration of 50\% and a wind fraction of 94\%. Other relevant research on the advantages of grid extensions for the integration of renewables, including reduced variability and smaller forecast errors, can be found in \cite{Tradewind,WE:WE410,Papae,Schaber,Schaber2,Egerer,Brown}.

In this paper the consequences of moving from a homogeneous spatial distribution of \gls{vres} and a uniform wind-to-solar mixing factor to a cost-optimal placement of \gls{vres} capacities around Europe are explored. The distribution of \gls{vres} plants is determined by at least two considerations. The first consideration is the geographical variation of the \gls{vres} quality. The resource quality is quantified through the capacity factor (CF) defined as
%-------------------
\begin{equation} \label{eq:1}
   \mbox{CF}  =  \frac{\mbox{average generation}}{\mbox{rated capacity}} .
\end{equation}
%-------------------
The capacity factor is a number between 0 and 1, where 0 means no generation and 1 means maximum generation at all times. Capacity factors for the European countries for onshore wind and solar PV are \highlight{calculated using \eqref{eq:1} and} listed in Table \ref{tab:capacity-factors}. The second consideration is the geographical variation of the temporal generation pattern for a given \gls{vres} type. This effect is particularly important for wind since Europe is large compared to the correlation length of wind of $\approx$ 600 km \cite{npg-15-803-2008,Widen2011,1748-9326-10-4-044004}, and wind therefore benefits from smoothing effects across the continent.

With these points in mind, the optimal heterogeneous spatial layouts of wind and solar PV across Europe is investigated and compared to the homogeneous layouts. The main point of comparison is the average cost of electricity, which is composed of the \gls{vres}, backup and transmission costs. Different approaches to cope with the resulting large number of degrees of freedom are considered. In the literature a common approach for heterogeneous systems is to use linear programming to optimise generation and transmission capacities simultaneously \cite{Czisch,Scholz,Hagspiel,Schlachtberger2016}, but this has the drawback that only a selection of representative weather conditions can be considered before computation times become infeasible. This makes the results susceptible to over-tuning to the weather selection. Other groups have used genetic algorithms to optimise generation, storage and transmission over a full year in Australia \cite{Elliston} and over three years in Europe, the Middle East and North Africa \cite{Bussar201440}. In this paper a novel local search algorithm was found to be most effective given the size and non-linear formulation of the optimisation problem, allowing 8 years of hourly weather to be considered.

A downside of pure optimisation approaches is that one loses an understanding of why particular solutions are optimal. This makes it hard to justify investment strategies to policy makers and to the public. To counter this downside,  more intuitive heuristic methods are developed here to construct layouts based on knowledge of resource quality, which are then compared to layouts obtained through optimisation. Distributions proportional to capacity factors (similar to the approach in \cite{Schaber2}) and distributions based on optimal portfolio theory that reduce risk, or standard deviation, of the in-feed (similar to approaches in \cite{HOL08,ROQ10,ROM11,THO16}) are considered and compared.

This paper is organised as follows: \Cref{sec:two} discusses the general modelling of the simplified European electricity system and the key infrastructure measures. \Cref{sec:three} describes the construction of heterogeneous layouts. In \Cref{sec:results} the performance of the different layouts and the resulting renewable penetrations for individual European countries are discussed. \Cref{sec:sensitivity-analysis} contains an analysis of the sensitivity of the results to variations in component costs. We conclude the paper with a discussion on the results and an outlook on future research.

\begin{table}[t]
\caption{Nomenclature}
\label{tab:nomenclature}
\centering
\begin{tabular}{ll}
\toprule
Name & Description\\
\midrule
$N$ & Set of nodes\\
$n,m$ & Node index\\
$l$ & Link index\\
$\Delta_n$ & Mismatch (VRES generation minus load)\\
$\alpha_n$ & Wind/solar mix\\
$\gamma_n$ & Renewable penetration\\
$G_n^{\{W,S,B\}}$ & Generation of wind, solar or backup\\
$G_n^R$ & Total renewable generation\\
$L_n$ & Load\\
$P_n$ & Net power balance\\
$\mathcal{K}_n^{\{W,S,B\}}$ & Wind, solar or backup capacity\\
$\mathcal{K}_l^T$ & Transmission capacity for link $l$\\
$E^B$ & Backup energy\\
$C_n$ & Curtailment\\
$B_n$ & Nodal balancing\\
$H$ & PTDF matrix\\
$F_l$ & Power flow on link $l$\\
CF$^{\{W,S\}}$ & Wind/solar capacity factor\\
$\langle x \rangle$ & Average value of $x$\\
$q$ & Quantile\\
$K$ & Heterogeneity parameter \\
\bottomrule
\end{tabular}
\end{table}

%*************************************************************************
\section{Methods I: general modeling}                      \label{sec:two}
%*************************************************************************

%***************************************************
\subsection{Renewable resource assessment}

Realistic time series describing the country-specific wind and solar PV power generation and the load are the starting point of the advocated weather-driven modelling of a simplified networked European electricity system. The utilized data set has been released from the Fraunhofer Institute for Wind Energy and Energy System Technology (formerly ISET, now IWES)~\cite{bofinger}. This data set covers the eight-year period from January 2000 to December 2007, has a temporal resolution of one hour and a spatial resolution of ${50\times 50}$~km$^2$ over all of Europe. Fixed country-specific capacity layouts have been used to first convert the weather data into onshore wind and solar power generation, and then to aggregate the latter over each of the 30 European countries; off-shore wind power generation is not considered. The country-specific load time series have been obtained from publicly available sources, extrapolated to cover missing data, and detrended from an annual growth of around 2\% to their year 2007 values. For more specific details see~\cite{bofinger,Heide2010}. A good alternative description of the conversion modelling is given in \cite{AND15,STA16,PFE16}.

\highlight{
The obtained wind and solar PV power generation time series have been rescaled to the capacity factors (CFs) from 2014. The latter have been determined in accordance with \cref{eq:1} from the EuroStat data for the installed capacities and the total generation for the year 2014~\cite{eurostat1,eurostat2,barometer}. The resulting CFs for each country and each technology are listed in Table~\ref{tab:capacity-factors}. For some of the countries (particularly smaller countries) no data was available or the calculated result was too uncertain because of too little or no installed capacity. For these countries the CFs are calculated as an average value from surrounding countries. These cases are marked by a star. Some countries with an already high installed capacity have a relatively low capacity factor compared to validated results from \cite{AND15}. For them the CFs have been raised by a small factor: 8\% in Germany, 4\% for wind in Spain, 4\% for solar in Italy and 2\% for wind in Great Britain. The final capacity factors presented in Table~\ref{tab:capacity-factors} are in accordance with \cite{STA16,PFE16}, which presents a critical assessment of current and future national capacity factors. Capacity factors for wind are likely to rise further in the future because of re-powering of wind turbines with more efficient, modern turbines at higher hub heights \cite{irena2}.}

\begin{table*}[t!]
\caption{
Capacity factors $\text{CF}_n^W$ and  $\text{CF}_n^S$ for onshore wind and solar PV for the European countries, derived from the EuroStat data~\cite{eurostat1,eurostat2,barometer}. The countries are sorted by their respective mean load $\langle L \rangle$ (in units of GW) over the \highlight{2000-2007} time series. \highlight{*: estimated values, see text for details.}
}  \label{tab:capacity-factors}
\begin{adjustbox}{center}
  {\setlength{\tabcolsep}{0.8em}
    \begin{tabular}{lrll|lrll|lrll}  \toprule
    &\textbf{$\langle L \rangle$} &\textbf{CF$^W_n$}&\textbf{CF$^S_n$}&
    &\textbf{$\langle L \rangle$} &\textbf{CF$^W_n$}&\textbf{CF$^S_n$}&
    &\textbf{$\langle L \rangle$} &\textbf{CF$^W_n$}&\textbf{CF$^S_n$} \\  \midrule
    \textbf{DE} & 54.2 & 0.18 & 0.12 & \textbf{FI} & 9.0 & 0.20 & 0.08 & \textbf{RS} & 3.9 & 0.20$^{*}$ & 0.14$^{*}$ \\
    \textbf{FR} & 51.1 & 0.22 & 0.12 & \textbf{CZ} & 6.6 & 0.19 & 0.11 & \textbf{IE} & 3.2 & 0.27 & 0.09$^{*}$ \\
    \textbf{GB} & 38.5 & 0.29 & 0.09 & \textbf{AT} & 5.8 & 0.20 & 0.11 & \textbf{BA} & 3.1 & 0.22$^{*}$ & 0.14$^{*}$ \\
    \textbf{IT} & 34.5 & 0.20 & 0.14 & \textbf{GR} & 5.8 & 0.21 & 0.17 & \textbf{SK} & 3.1 & 0.21$^{*}$ & 0.12 \\
    \textbf{ES} & 24.3 & 0.26 & 0.21 & \textbf{RO} & 5.4 & 0.21 & 0.14 & \textbf{HR} & 1.6 & 0.25 & 0.14$^{*}$ \\
    \textbf{SE} & 16.6 & 0.25 & 0.09 & \textbf{BG} & 5.1 & 0.22 & 0.14 & \textbf{LT} & 1.5 & 0.25 & 0.12$^{*}$ \\
    \textbf{PL} & 15.2 & 0.23 & 0.12$^{*}$ & \textbf{PT} & 4.8 & 0.28 & 0.18 & \textbf{EE} & 1.5 & 0.20 & 0.10$^{*}$ \\
    \textbf{NO} & 13.7 & 0.29 & 0.09$^{*}$ & \textbf{CH} & 4.8 & 0.20$^{*}$ & 0.11$^{*}$ & \textbf{SI} & 1.4 & 0.20$^{*}$ & 0.13 \\
    \textbf{NL} & 11.5 & 0.23 & 0.09 & \textbf{HU} & 4.4 & 0.22 & 0.13$^{*}$ & \textbf{LV} & 0.7 & 0.23 & 0.11$^{*}$ \\
    \textbf{BE} & 9.5 & 0.27 & 0.11 & \textbf{DK} & 3.9 & 0.31 & 0.10 & \textbf{LU} & 0.7 & 0.16 & 0.10 \\ \bottomrule
    \end{tabular}
  }
\end{adjustbox}
\end{table*}

%**************************************
\subsection{The electricity network}
\label{subsec:network}

The European electricity network is modelled as a simplified 30-node model, where each node represents a country. For each node $n$ the generation from \gls{vres} \highlight{(see Table \ref{tab:nomenclature} for a summary of nomenclature)},
% -------------------
\begin{equation}
  G^{R}_{n}(t) = G_{n}^{W}(t) + G_{n}^{S}(t),
\end{equation}
% -------------------
can be expressed through two parameters. The penetration $\gamma$ determines the amount of renewable energy generated relative to the mean load of the node,
% -------------------
\begin{equation}
  \mean{G^{R}_{n}} = \gamma_{n} \mean{L_{n}} ,
\end{equation}
% -------------------
while the mixing parameter $\alpha$ fixes the wind-to-solar ratio,
%-------------------
\begin{align}
  \mean{G^{W}_{n}} &=  \alpha_{n} \mean{G_{n}^{R}}  , \\
  \mean{G^{S}_{n}} &=  \paren{1- \alpha_{n}} \mean{G_{n}^{R}}  .
\end{align}
%-------------------
Other forms of renewable power generation are neglected in this simplistic modelling approach.

The nodal difference between \gls{vres} generation and load
%-------------------
\begin{equation} \label{eq:deltan}
  \Delta_{n}(t) = G^{R}_{n}(t) - L_{n}(t)
\end{equation}
%-------------------
is called the mismatch. To avoid power outages, the demand must be met at all times. Since storage is not considered, any power deficits must be covered by backup generation. Dispatchable resources are not modelled explicitly, but are considered as part of the backup generation. If $\Delta_{n}(t) \geq 0$, excess energy $C_{n}(t)$ must be curtailed, while if $\Delta_{n}(t) < 0$ backup generation $G^{B}_{n}(t)$ is needed. Together the two terms form the nodal balancing $B_{n}(t) = C_{n}(t) - G^{B}_{n}(t)$. It is possible to lower the balancing needs with transmission. Nodes with excess generation export energy $E_{n}(t)$, allowing nodes with an energy deficit to import energy $I_{n}(t)$ to (partly) cover their energy deficit. The nodal injection, $E_{n}(t) - I_{n}(t)$, is denoted $P_{n}(t)$. This leads to the nodal balancing equation,
% -------------------
\begin{equation}  \label{eq:nodal-balancing}
  G^{R}_{n}(t) - L_{n}(t) = B_{n}(t) + P_{n}(t) \; ,
\end{equation}
% -------------------
The vector of nodal injections is called the injection pattern, and fullfills $\sum_n P_n(t) = 0$. The actual imports and exports, and thus the injection pattern, depend on the dispatch of the nodal balancing. The synchronised balancing scheme,
% -------------------
\begin{equation}  \label{eq:synch}
  B_n(t)  =  \frac{ \langle L_n \rangle }{ \sum_k \langle L_k \rangle }  \sum_m \Delta_m(t) \; ,
\end{equation}
% -------------------
where all nodes are curtailing/generating backup syn\-chron\-ously (relative to $\mean{L_{n}}$), fulfills two top priorities: it minimises the total backup generation for each time step and it minimises the overall backup capacity \cite{Rolando2015}. This stylised synchronised balancing scheme has also been chosen in view of the layout optimisation, since the computational time for an update step is much smaller than for other dispatch schemes, like for example the localised flow scheme used in two previous publications \cite{Rodriguez2014,sarah}.

The injection pattern is fixed by Eqs.\ (\ref{eq:nodal-balancing}) and (\ref{eq:synch}), and determines the power flows on the links $l$:
% -------------------
\begin{equation}  \label{eq:FHP}
  F_{l}(t) = \sum_n H_{ln} P_n(t)  \; .
\end{equation}
% -------------------
The linear relationship follows from the DC approximation, which is known to be a good approximation for high-voltage flows. For the
Power Transfer Distribution Factors $H_{ln}$ we have assumed unit susceptances \cite{Rolando2015}, allowing its construction from the Moore-Penrose pseudo inverse of the underlying network Laplacian.

%****************************************
\subsection{Infrastructure measures}
\label{sec:key}

Following \cite{Sensitivity}, the energy system cost is calculated based on a few key measures. Besides the cost of the \gls{vres} capacities, $\mathcal{K^{W}}$ and $\mathcal{K^S}$, costs for the backup system and the transmission network are included. The backup system cost is split into two components, the cost of backup capacity $\mathcal{K}^{B}$ and the cost of backup energy $E^{B}$. The backup capacity cost covers expenses related to construction and to keeping the power plants online while the backup energy cost accounts for actual fuel costs. Expressed in units of the average annual load, the backup energy is given by
% -------------------
\begin{equation}
  \label{eq:backup-energy}
  E^{B}  =  \frac{\sum_{n} \sum_{t} G^{B}_{n}(t)}{\sum_{m} \sum_{t} L_{m}(t)}
             =  \frac{\sum_{n} \langle G^{B}_{n} \rangle}{\sum_{m} \langle L_{m} \rangle} \; .
\end{equation}
% -------------------
In principle, the backup capacity is fixed by a single extreme event. However with this definition, the results will be highly coupled to the particular data set used. To decrease the coupling, the 99\% quantile is used rather than the maximum value,
% -------------------
\begin{equation}
  \label{eq:2}
  q_{n} = \int _{0} ^{K_{n}^{B}} p_{n}(G^{B}_{n})\ dG^{B}_{n} \; ,
\end{equation}
% -------------------
where $p_{n}(G^{B}_{n})$ is the time sampled distribution of backup generation and $q_{n} = 0.99$. With this choice, the backup system will be able to fully cover the demand 99\% of the time. The remaining 1\% is assumed to be covered by unmodelled balancing initiatives, e.g. demand side management. Given the nodal values $\mathcal{K}^{B}_{n}$, the overall backup capacity
% -------------------
\begin{equation}
  \mathcal{K}^{B} = \sum_{n} \mathcal{K}^{B}_{n}
\end{equation}
% -------------------
is calculated by summation.

In analogy, the transmission capacity $\mathcal{K}^{T}_{l}$ is defined so that the flow is met 99\% of the time. Transmission can be positive and negative, but since links are assumed bidirectional, only the magnitude (not the sign) of the flow is to be considered. Hence
% -------------------
\begin{equation}
  \label{eq:link-cap}
  q_{l}  =  \int_{0}^{K_{l}^{T}} p_{l}(|F_{l}|)\ d|F_{l}| \; ,
\end{equation}
% -------------------
where $p_{l}(|F_{l}|)$ is the time sampled distribution of absolute flows and $q_{l} = 0.99$. Since the link length varies, $\mathcal{K}^{T}$ is not calculated directly by summation, but instead as a weighted sum,
% -------------------
\begin{equation}
  \mathcal{K}^{T} = \sum_{l} \mathcal{K}^{T}_{l} d_{l},
\end{equation}
% -------------------
where $d_{l}$ denotes the length of link $l$. Link lengths are estimated as the distance between the country capitals.

In this paper $E^{B}$ will be expressed in units of average annual load, $\mathcal{K}^{B}$ in units of average hourly load and $\mathcal{K}^{T}$ in units of average hourly load $\times$ megametre.

%******************************
\subsection{Cost modelling}
\label{sec:cost-modelling}

Cost assumptions for the elements of an electricity system vary greatly across the literature. In this study, the cost assumptions published by \cite{Sensitivity} have been adapted with a single modification. The cost of solar has been reduced by 50\% in accordance with near future solar PV panel price projections \cite{irena}. The resulting estimates are listed in Table \ref{tab:cost-assumptions}. In general, the cost assumptions are in the low end for \gls{vres} which reflects the expectation that the cost of \gls{vres} will go down in the future as the penetration increases.  Backup generation is priced based on the cost of Combined Cycle Gas Turbines (CCGTs).

% -------------------
\begin{table}[t!]
\centering
\caption{
Cost assumptions for different assets separated into \gls{CapEx} and fixed/variable \gls{OpEx} together with their expected life times.
}  \label{tab:cost-assumptions}
  {\setlength{\tabcolsep}{0.2em}
  \begin{tabular}{lrrrr}  \toprule
    \textbf{Asset} & \textbf{\gls{CapEx} }& \textbf{\gls{OpEx}$_{\text{fixed}}$} & \textbf{\gls{OpEx}$_{\text{var}}$} & \textbf{Life time}\\
    & [\euro/W] & [\euro/kW/y] & [\euro/MWh] & [years]\\ \midrule
    CCGT & 0.90 & 4.5 & 56.0 & 30\\
    Solar PV & 0.75 & 8.5 & 0.0 & 25\\
    %Offshore wind & 2.00 & 55.0 & 0.0 & 20\\
    Onshore wind & 1.00 & 15.0 & 0.0 & 25\\
    \bottomrule
  \end{tabular}}
\end{table}
% -------------------

From the \gls{vres} penetration, the mixing factor and the mean load, the mean generation of each node can be calculated. Dividing by the associated capacity factor, the capacity is obtained. Except for transmission capacity, the present value of each element can be calculated directly as
% -------------------
\begin{equation}
  \label{eq:6}
  V = \text{\gls{CapEx}} + \sum_{t=1}^{T_\mathrm{life}} \frac{\text{\gls{OpEx}}_{t}}{\paren{1 + r}^{t}} \; ,
\end{equation}
% -------------------
where $r$ is the rate of return assumed to be 4\% per year. The transmission capacity cannot be translated directly into cost as the cost depends on the length and the type of the link. Link costs are assumed to be 400\euro{} per km per MW for AC links and 1,500\euro{} per km per MW for HVDC links. For HVDC links, an additional cost of 150,000\euro{} per MW per converter station pair (one at each end) is added \cite{McKinsey, Schaber, Schaber2}. The layout of AC and HVDC lines has been constructed by \cite{Rodriguez2014} according to the existing European network reported by ENTSO-E for the year 2011 \cite{ENTSOEcaps} and new predicted lines until 2014 \cite{NorNed, BritNed}. It is shown in Figure \ref{fig:links}.

To allow for comparison of different system layouts, the \gls{lcoe} is a convenient measure \cite{Sensitivity,UEC13,HIR15}. The \gls{lcoe} is the cost that every generated unit of energy consumed during the lifetime of the project has to match the present value of investment \cite{Short1995},
% -------------------
\begin{equation}
  \label{eq:7}
  \text{LCOE}_V = \frac{V}{\sum_{t=1}^{T_\mathrm{life}} \frac{L_{EU,t}}{\paren{1+r}^{t}}} .
\end{equation}
% -------------------
Since the life time of the system elements differs, the \gls{lcoe} is evaluated separately for each system element from each respective present value. The \gls{lcoe} for the complete system is calculated by summation. Life times of 25 years for solar PV and onshore wind, 30 years for CCGT plants and 40 years for transmission infrastructure were assumed.

%********************************************************************
\section{Methods II: heterogeneous layouts}  \label{sec:three}
%********************************************************************

The simplest way to distribute the renewable resources is to assign them homogeneously (relative to the mean load of the node) so that $\gamma_{n} = \gamma_{EU} = 1$ and $\alpha_{n} = \alpha_{EU}$. This homogenous layout is denoted as HOM. However this assignment might not be ideal since the capacity factors vary significantly between the nodes. Three heuristic schemes and a straightforward optimisation for the construction of heterogeneous layouts will be presented in the following four subsections. The naming of the distribution algorithms is summarised in Table \ref{tab:algorithms}.

%**************************************************************
\subsection{Heuristic layout I: CF proportional (CFprop)}
\label{sec:beta-layout}

An intuitive first approach, called CFprop, is to assign resources proportional to the CF, or more general to the CF raised to an exponent $\beta$. For a wind-only layout, the nodal renewable penetrations $\gamma_n$ are given by
% -------------------
\begin{equation}  \label{eq:8}
  \gamma_{n}^{W}  =  \frac{\paren{\text{CF}^{W}_{n}}^{\beta} \langle L_{EU} \rangle}
                                          {\sum_{m} \paren{\text{CF}^{W}_{m}}^{\beta} \langle L_{m} \rangle}
                                   \gamma_{EU}  \; ,
\end{equation}
% -------------------
where $\gamma_{EU}$ is the overall penetration assumed to be 1. An equivalent expression for the solar-only layout is obtained by the substitution $W \to S$. Examples for $\beta$ = 1 are shown in \Cref{fig:examples}a for the wind- and solar-only layouts. In the layout illustrations, each bar represents a country.

CFprop layouts for any value of $\alpha$ can be constructed as a linear combination of the wind and solar only layouts with
% -------------------
\begin{equation}  \label{eq:9a}
  \gamma_{n} = \alpha_{EU} \gamma^{W}_{n} + (1-\alpha_{EU}) \gamma^{S}_{n}
\end{equation}
% -------------------
and
% -------------------
\begin{equation} \label{eq:9b}
  \alpha_{n} = \frac{\alpha_{EU} \gamma_{n}^{W}}{\alpha_{EU} \gamma_{n}^{W} + (1-\alpha_{EU}) \gamma_{n}^{S}}  \; .
\end{equation}
% -------------------

For practical reasons, it is not possible to realise extremely heterogeneous layouts. On the one hand the geographical potentials for \gls{vres} installations in countries with good renewable resources may be a limiting factor. On the other hand countries with poor renewable resources may not want to become too dependent on imports. To constrain heterogeneity, the heterogeneity parameter K is introduced by requiring
% -------------------
\begin{equation}
  \label{eq:k-factor}
  \frac{1}{\text{K}} \leq \gamma_{n} \leq \text{K} .
\end{equation}
% -------------------
With this definition, K = 1 corresponds to a homogeneous layout while K = $\infty$ represents unconstrained heterogeneity. For the CFprop layouts, each value of K translates into an $\alpha$-dependent value of $\beta$. For a given value of $\alpha$, the corresponding $\beta$ value is found by increasing $\beta$ until the first country violates \cref{eq:k-factor}. At the mix $\alpha = 0.86$ the values $K= 1, 2, 3$ correspond to $\beta = 0.00, 1.92, 2.91$, respectively.

%***********************************************************************
\subsection{Heuristic layout II: extreme K-constrained (CFmax)}
\label{sec:CF-layout}

\begin{figure}[t]
\centering
  \begin{subfigure}{\columnwidth}
    \includegraphics[width = \chromowidth, center]{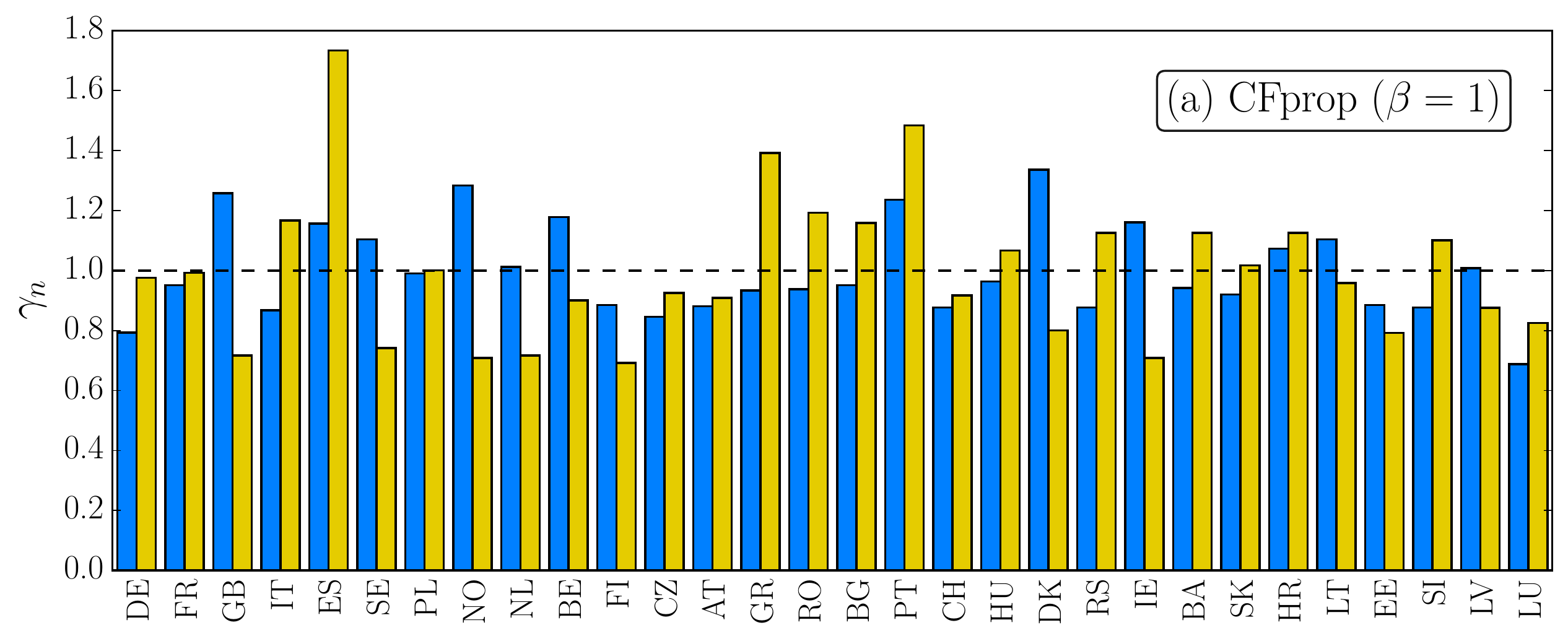}
    \vspace{-1em}
    \label{fig:betaExamples}
  \end{subfigure}
  \begin{subfigure}{\columnwidth}
    \includegraphics[width = \chromowidth, center]{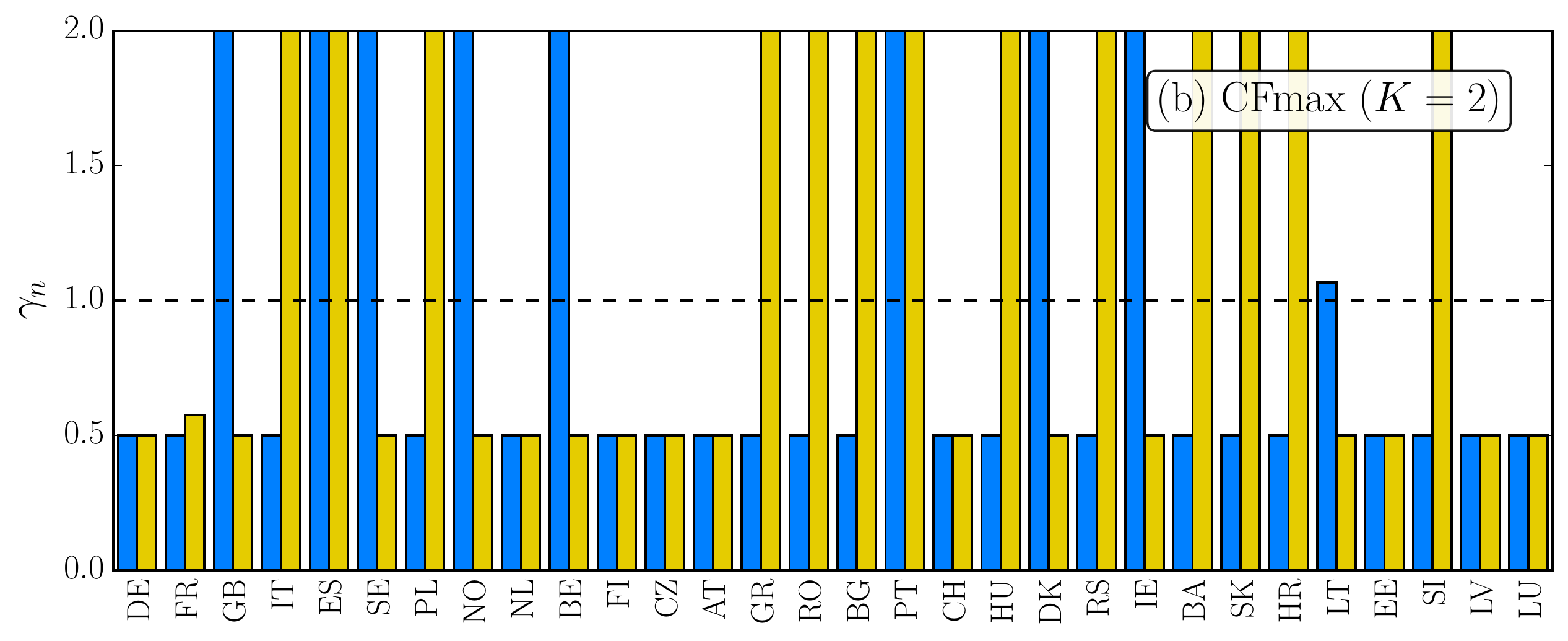}
    \vspace{-1em}
    \label{fig:cfMaxExamples}
  \end{subfigure}
  \begin{subfigure}{\columnwidth}
    \includegraphics[width = \chromowidth, center]{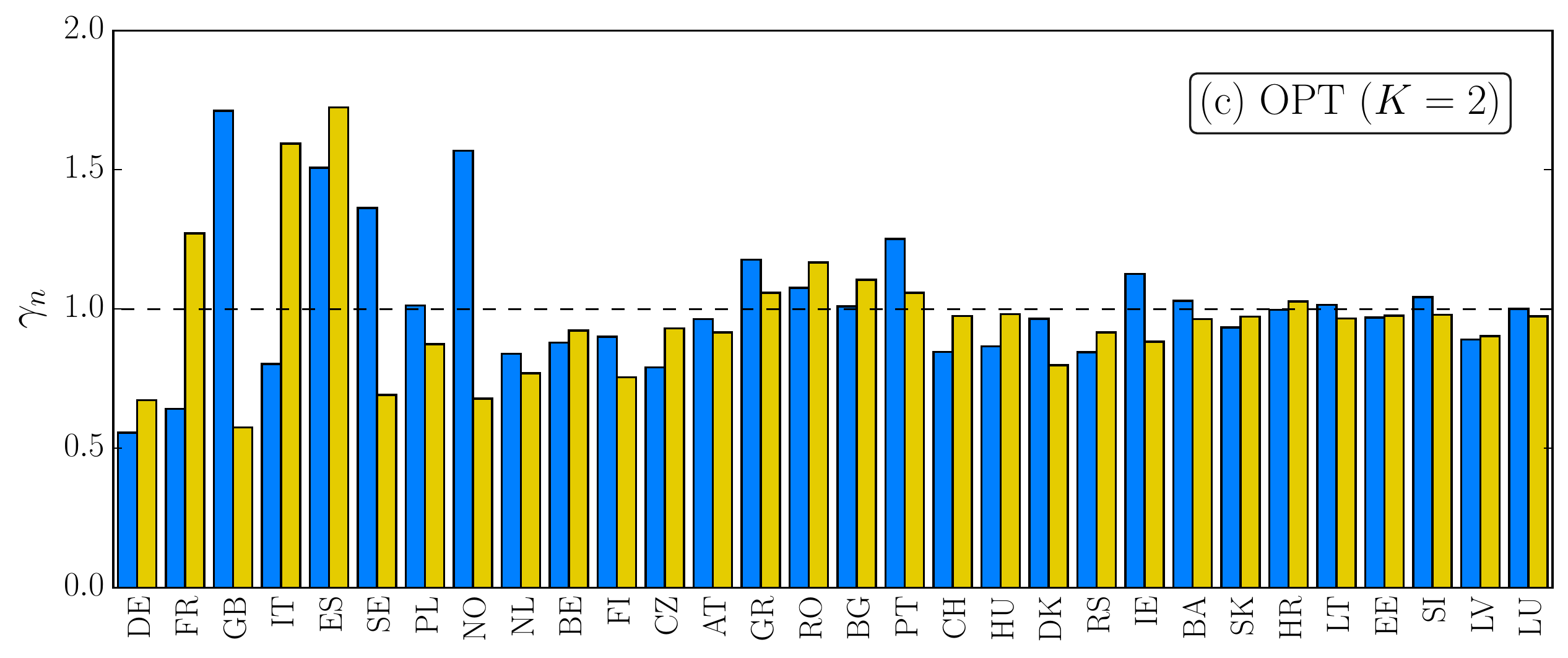}
    \vspace{-1em}
  \end{subfigure}
\caption{
Examples of heuristic (blue) wind-only and (yellow) solar-only layouts: (a) CFprop with $\beta=1$, (b) CFmax constrained to $K=2$, and (c) Pareto optimal OPT layouts obtained with $K=2$.
}  \label{fig:examples}
\end{figure}

Although the overall capacity factor of a CFprop layout for $\beta$ $>$ 0 is higher than the capacity factor of the homogeneous layout, it is possible to achieve an even higher capacity factor without violating the constraints in \cref{eq:k-factor}. In the wind- and solar-only cases, the capacity factor is maximised by assigning $\gamma_{n}$ = K to the countries with the highest capacity factor and $\gamma_{n} = \frac{1}{\text{K}}$ to the remaining countries, except for a single in-between country which is fixed by the constraint
% -------------------
\begin{equation}  \label{eq:norm}
  \sum_{n} \gamma_{n} \mean{L_{n}} = \mean{L_{EU}}.
\end{equation}
% -------------------
The wind- and solar-only cases of the CFmax layout constrained by $K=2$ are shown in \Cref{fig:examples}b. Similar to the CFprop layouts, the CFmax layouts for arbitrary $\alpha_{EU}$ values can be constructed as linear combinations (\ref{eq:9a}) of the wind- and solar-only layouts.

% -------------------
% -------------------

%****************************************
\subsection{Heuristic layout III: OPT}
\label{sec:OPT-layout}

The optimal portfolio theory (OPT) is well known in mathematical finance \cite{MAR52}. It discusses different assets obtained from the tradeoff between maximizing their return and minimizing their risk. This concept has also been applied to find optimal deployment of wind and solar energy resources in large-scale energy systems \cite{ROQ10,ROM11,THO16}, where the overall capacity factor has been treated as the return and the variance of the renewable power generation as the risk. In modified form, we will use OPT to further explore \gls{vres} capacity layouts over Europe with low system cost of electricity.

The overall capacity factor of a wind-only ($\gamma_n^W$) or solar-only ($\gamma_n^S$) layout is defined as
% -------------------
\begin{equation} \label{eq:CFEU}
   CF^{W/S}_{EU}  =  \frac{\mean{L_{EU}}}{\mathcal{K}^{W/S}_{EU}}  \; ,
\end{equation}
% -------------------
where
% -------------------
\begin{equation}
   \mathcal{K}^{W/S}_{EU}  =  \sum_n \frac{\gamma_n^{W/S}\mean{L_n}}{CF_n^{W/S}}
\end{equation}
% -------------------
represents the overall installed capacity. The overall capacity factor is a useful measure of return as this is of high importance for investors of renewable generation capacity. Investors seek to minimise the overall capacity investment, which corresponds to maximising the overall capacity factor.

OPT's second measure is risk, for which we select the relative standard deviation
$\sigma_{\Delta}/\mean{L_{EU}}$ of the overall mismatch
% -------------------
\begin{equation}  \label{eq:DeltaEU}
   \Delta_{EU}(t)  =  \sum_n \Delta_n(t)
\end{equation}
% -------------------
based on the country-specific mismatches (\ref{eq:deltan}). The smaller the risky standard deviation becomes the more likely is the reduced need for a backup infrastructure, which an investor tries to minimise \cite{HOL08}.

A possible heterogeneous wind- or solar-only capacity layout is sampled from a Monte Carlo procedure. The country-specific renewable penetrations $\gamma_n$ are randomly and independently drawn from a Beta distribution
% -------------------
\begin{eqnarray} \label{eq:Beta}
    p(\gamma)
      &=&  \frac{\Gamma(\beta_1+\beta_2)}{\Gamma(\beta_1)\Gamma(\beta_2)}
               \left( \frac{K}{K^2-1} \right)^{\beta_1+\beta_2-1}
               \nonumber \\
      & &   \left( \gamma - \frac{1}{K} \right)^{\beta_1-1}
               \left( K - \gamma \right)^{\beta_2-1}
\end{eqnarray}
% -------------------
defined on the compact support (\ref{eq:k-factor}). $\Gamma(\beta)$ is the Gamma function. The two shape parameters $\beta_1$ and $\beta_2$ are determined by requiring $\mean{\gamma_n}=1$ and by envoking the maximum entropy principle \cite{HON98} to maximal smear out the Beta distribution over the interval (\ref{eq:k-factor}). For $K=2$ the two parameters result in $\beta_1=0.80$, $\beta_2=1.61$, and for $K=3$ they are $\beta_1=0.86$, $\beta_2=2.57$.  A capacity layout sampled with this procedure does not necessarily meet the requirement \eqref{eq:norm}. For such cases, all $\gamma_n$ are uniformly rescaled upwards or downwards until the requirement is fulfilled. During the rescaling some of the penetration parameters hit the $K$ constraint \eqref{eq:k-factor}, and are then frozen for the remainder of the rescaling procedure.

\begin{figure}[t] \centering
\includegraphics[width = \columnwidth]{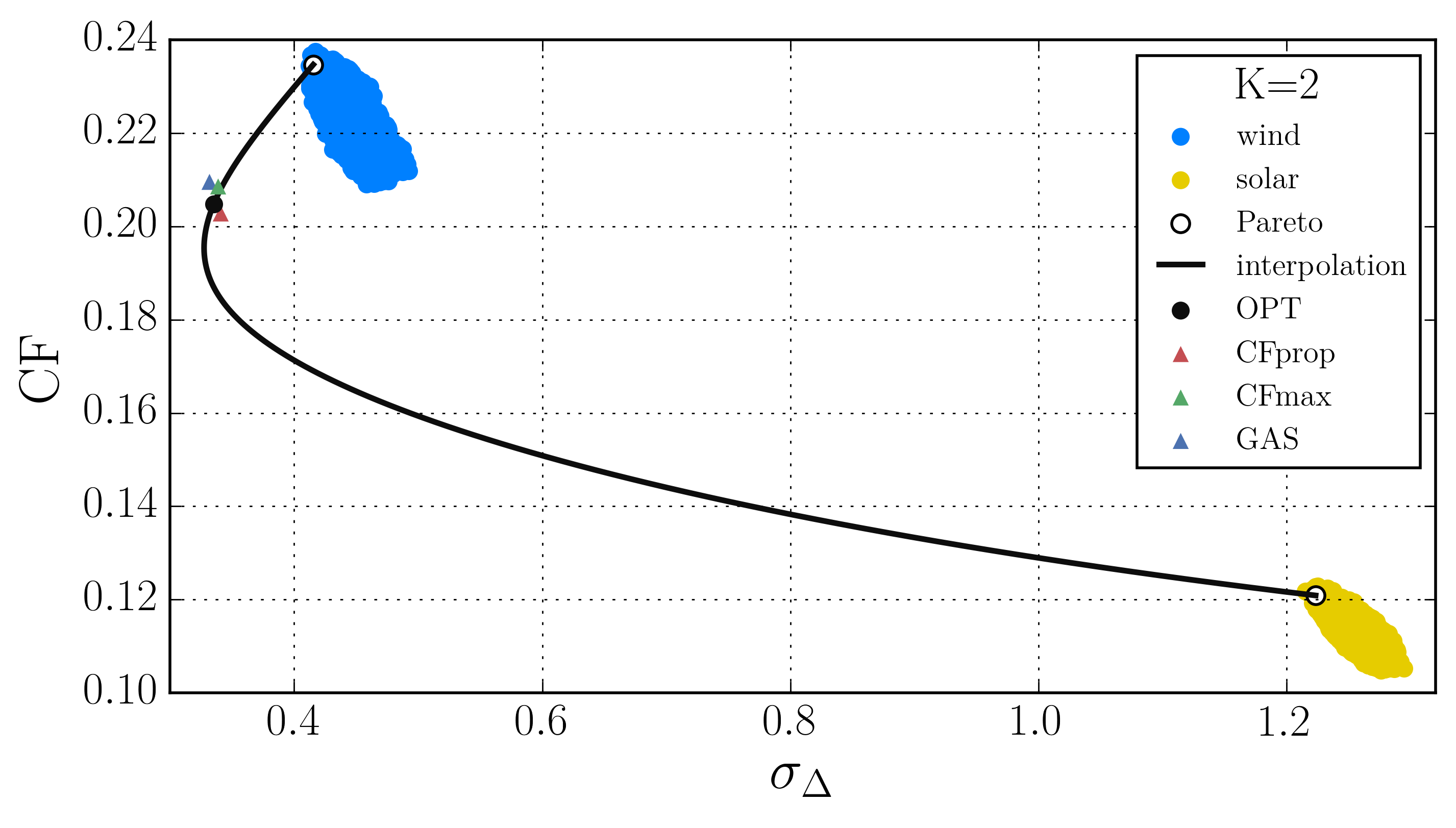}%
\caption{
Scatter clouds for (blue) wind-only and (yellow) solar-only capacity layouts. The diagram plots the overall capacity factor (\ref{eq:CFEU}) vs.\ the standard deviation of the overall mismatch  (\ref{eq:DeltaEU}). The distribution (\ref{eq:Beta}) and the constraint (\ref{eq:k-factor}) with $K=2$ have been used for the Monte Carlo simulations. The white point in the upper left cloud corners indicate Pareto optimal layouts; see also Figure \ref{fig:examples}c. The line connecting the wind- and solar-only Pareto optimal layouts results from the interpolation \highlight{between these layouts in} (\ref{eq:9a}). The black point marked on this line represents the OPT layout with minimum LCOE. For comparison, the three triangle points mark the (orange) optimal CFprop, (green) optimal CFmax and (blue) optimised GAS layouts for $K=2$.
}  \label{fig:portfolios}
\end{figure}

The wind-only and solar-only portfolios for $K=2$ are shown in \fref{fig:portfolios} in blue and yellow, respectively, with the overall mismatch measure on the first axis and the overall capacity factor measure on the second axis. Each of the portfolios consists of 100000 layouts. Due to the elongated shape of the portfolios there is no clear extended Pareto front in the upper left corners. The Pareto front defines a line, for which at the same time the standard deviation of the overall mismatch (risk) can not be reduced further for a fixed overall capacity factor and the overall capacity factor (return) can not be increased further for a fixed standard deviation of the overall mismatch. For both portfolios we identify a single point to characterise minimum risk and maximum return. This is done by extracting a subset of the points which are simultaneously a part of the top 200 capacity factors and bottom 200 standard deviations. For $K=2$ this leaves a sample of 28 layouts for wind and 67 layouts for solar to average and to calculate the respective new overall capacity factor and new overall standard deviation. The resulting points are plotted in white on top of the portfolios. The layouts of these two Pareto optimal points are shown in \fref{fig:examples}c.

In order to find an optimal combined layout, we interpolate between the Pareto-optimal wind-only and solar-only layouts according to (\ref{eq:9a}) and (\ref{eq:9b}). This interpolation conserves the constraint (\ref{eq:norm}) and results in the line shown in \fref{fig:portfolios}. Apparently some of the interpolated layouts are able to reduce the standard deviation of the global mismatch further. The interpolated layout marked with a black dot comes with the mixing parameter $\alpha_{EU}=0.87$.

%*********************************
\subsection{Optimised layouts}
\label{sec:optimized-layouts}

\begin{algorithm*}
\caption{
Pseudo code for the greedy axial search (GAS) routine. The \textit{Evaluate} function calculates the associated cost of each new solution, and all new solutions are thereupon sorted by the \textit{Sort} function in ascending order.
}  \label{gasAlgo}
\begin{algorithmic}
    \Function{GreedyAxialSearch}{}
    \State \textit{best} $\gets$ solution selected randomly from within the
    solution space
    \State\textit{deltaCost} $\gets$ $\infty$
    \State\textit{stepSize} $\gets$ maxStepSize
    \While{\textit{stepSize} $>$ \textit{minStepSize}}
    \While{\textit{deltaCost} $>$ \textit{tolerance}}
    \For{index \textit{i} = 1 to 2N}
    \State \textit{trailSolutions[i]} $\gets$ StepUp(\textit{best},\textit{i},\textit{stepSize})
    \State \textit{trailSolutions[i+2N]} $\gets$ StepDown(\textit{best},\textit{i},\textit{stepSize})
    \EndFor
    \State Evaluate(\textit{trailSolutions})
    \State Sort(\textit{trailSolutions})
    \State \textit{deltaCost} $\gets$ cost of \textit{best} minus cost of \textit{trailSolutions[1]}
    \If {\textit{deltaCost} $>$ 0}
    \State \textit{best} $\gets$ \textit{trailSolutions[1]}
    \EndIf
    \EndWhile
    \State \textit{stepSize} $\gets$ \textit{stepSize}/2
    \EndWhile
    \State \Return \textit{best}
    \EndFunction
\end{algorithmic}
\end{algorithm*}

The full optimisation of the layouts is considered, with the objective to minimise the \gls{lcoe} with respect to the 60 variables $\gamma_{1}, ..., \gamma_{N}, \alpha_{1}, ..., \alpha_{N}$ for the $N=30$ countries. Given the high dimensionality of the search space, a number of optimisation algorithms were tested including the Nelder-Mead method \cite{nelder}, simulated annealing \cite{sa}, genetic algorithms \cite{ga} and cuckoo search \cite{cs}. It was found that the continuous enforcement of the normalisation criterion (\ref{eq:norm}) generally decreased the performance of the tested algorithms, and for that reason a new hybrid algorithm was developed to address this problem. While being a classical greedy algorithm in the sense that the locally optimal choice is always taken, the renormalisation problem was circumvented by moving only along the axial directions. The algorithm has been denoted Greedy Axial Search (GAS).

When a solution is renormalised, all $\gamma$ values are scaled either up or down. Therefore, it is possible that some $\gamma$ values end up outside the boundary (\ref{eq:k-factor}). The $\gamma$ values are fixed at the boundary and the rescaling is only applied to the remaining free $\gamma$ values. In general this approach is problematic since it can change the direction of the search. This is circumvented by holding the specific $\gamma$ value constant that is considered during the step up/down procedure along a given axis. In this way only some $\gamma$ values are scaled down/up and the feasibility of moving up/down along the considered axis can be determined. This is the underlying principle of \gls{gas}.

As any greedy algorithm, the \gls{gas} algorithm works by taking the locally optimal choice. Hence the feasibility for each direction is evaluated, but only the best choice is accepted. This process is repeated until a convergence criterion is fulfilled. At this point the step size is reduced and the iterative optimisation procedure repeated until the step size drops below some tolerance. The algorithm structure is sketched in Algorithm \ref{gasAlgo}. The \textit{StepUp} and \textit{StepDown} subroutines generate new solutions by stepping a solution (first argument) up/down along axis $i$ (second argument) with some step size (third argument) after which the solution is renormalised as described above. Values of $maxStepSize = 1$, $minStepSize = 5  \cdot 10^{-4}$ and $tolerance = 10^{-4}$ were found to be appropriate.

All optimised layouts have been obtained using the GAS routine. These layouts will be denoted GAS layouts.
Constraining the transmission and thereby reducing the transmission capacity can lead to an overall lower LCOE. This is discussed in Section~\ref{subsec:TransCap}. The layouts resulting from this additional optimisation will be denoted GAS* layouts.

\begin{table}[t!]  \centering
\caption{
Summary of the algorithms for distributing VRES.
}
\label{tab:algorithms}
\begin{tabular}{lp{6.5cm}}
    \toprule
Name & Brief description \\
    \midrule
    HOM & Homogeneous distribution proportional to the mean of each country's load\\
    CFprop & Distribution proportional to a power $(CF)^\beta$ of the capacity factor $CF$ \\
    CFmax & Assignment to each country $\gamma_n$ extremised within $\frac{1}{K} \leq \gamma_n \leq K$ depending on $CF$ \\
    OPT & Distribution using Optimal Portfolio Theory \\
    GAS & Distribution optimised using Greedy Axial Search algorithm \\
    GAS* & As GAS, but with optimally constrained transmission \\
    GASnoT & As GAS but with no transmission between countries, so that each country is self-sufficient at all times \\
    \bottomrule
\end{tabular}
\end{table}

%*****************************************
\section{Results}     \label{sec:results}
%*****************************************

The optimal heuristic layouts CFprop, CFmax, OPT as well as the optimized layouts GAS will be discussed in the next three subsections, first for $K=1$, then for $K=2$, and finally for $K=3$. The fourth subsection focuses on the transmission capacities.

%*****************************
\subsection{$K=1$ layouts}
\label{subsec:ResultsK1}

\begin{figure*}[t!]  \centering
  \includegraphics[trim={0.1cm 0.4cm 0.25cm 0.2cm},clip,width = \textwidth]{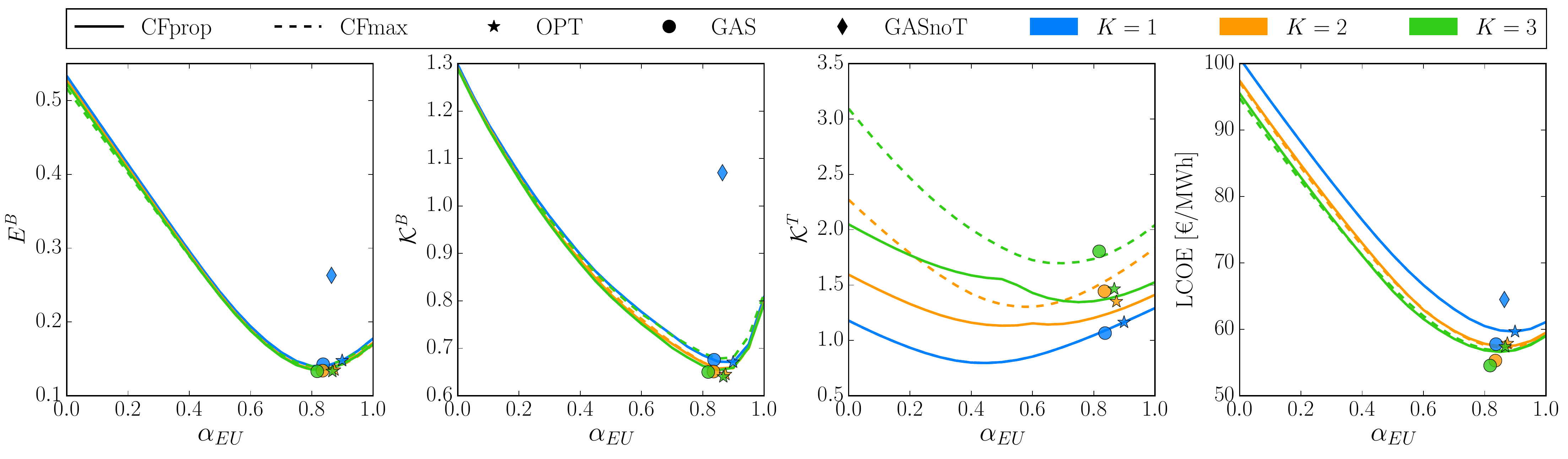}
\caption{
Overview of the infrastructure measures: (a) the backup energy $E^B$ (in units of average annual European load), (b) the backup capacity $\mathcal{K}^B$ (in units of average hourly European load), (c) the transmission capacity $\mathcal{K}^T$ (in units of average hourly European load times megametre) and (d) the associated \gls{lcoe} as a function of $\alpha_{EU}$. The CFprop and CFmax layouts are shown as solid and dashed lines respectively. The dependence of the OPT layouts on $\alpha_{EU}$ is not shown; only the interpolations leading to a LCOE minimum are plotted as asterisks. The GAS layouts are plotted as dots. The blue diamond represents the GASnoT layout. Different constraints are shown: K = 1 (blue), 2 (yellow) and 3 (green).
}  \label{fig:overview}
\end{figure*}

By construction, the layouts CFprop, CFmax and OPT become identical and homogeneous for $K=1$. Due to Eq.\ (\ref{eq:k-factor}), their respective renewable penetrations are $\gamma_n=1$. Moreover, according to Eq.\ (\ref{eq:9b}) their renewable mixes $\alpha_n = \alpha_{EU}$ also turn out to be independent of the country index. For these strictly homogeneous layouts Figure \ref{fig:overview} shows the dependence of the key infrastructure measures on $\alpha_{EU}$ as the blue curves. For the backup energy and backup capacity, the optimal mixing parameters are located around $\alpha_{EU} = 0.85$, which is slightly larger than the values found by \cite{Heide2010,Heide2011}. For the transmission capacity, the minimum occurs around $\alpha_{EU} = 0.45$. The main measure of interest, the \gls{lcoe}, has a minimum at $\alpha_{EU} = 0.90$. The high cost at $\alpha_{EU} = 0$ is caused by a combination of high backup energy/capacity costs and the fact that the CF of solar is generally lower than for onshore wind. The cost of producing one unit of energy is thus higher for solar than for onshore wind even though the specific \gls{CapEx} is lower for solar.

\begin{figure}[t]  \centering
  \begin{subfigure}{\columnwidth}
    \includegraphics[width = \chromowidth, center]{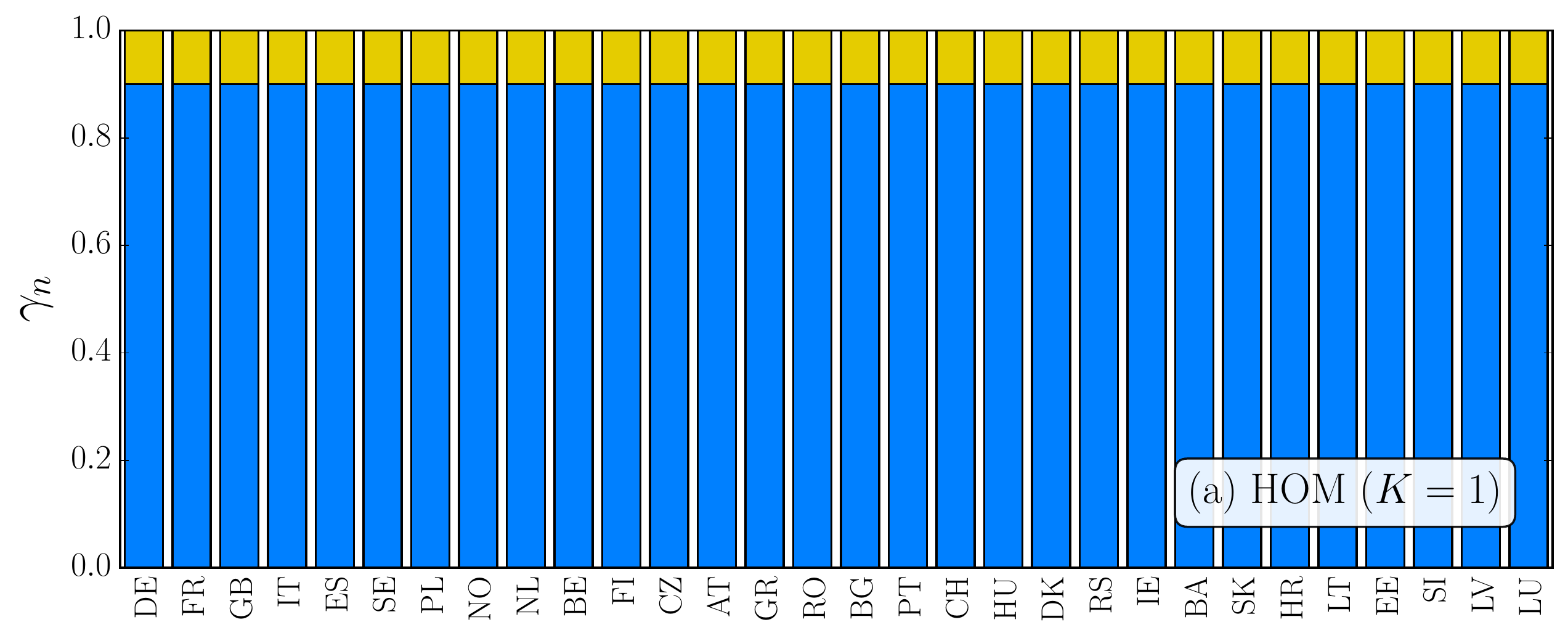}
    %\caption{Homogeneous layout.}
    \vspace{-1em}
  \end{subfigure}
  \begin{subfigure}{\columnwidth}
    \includegraphics[width = \chromowidth, center]{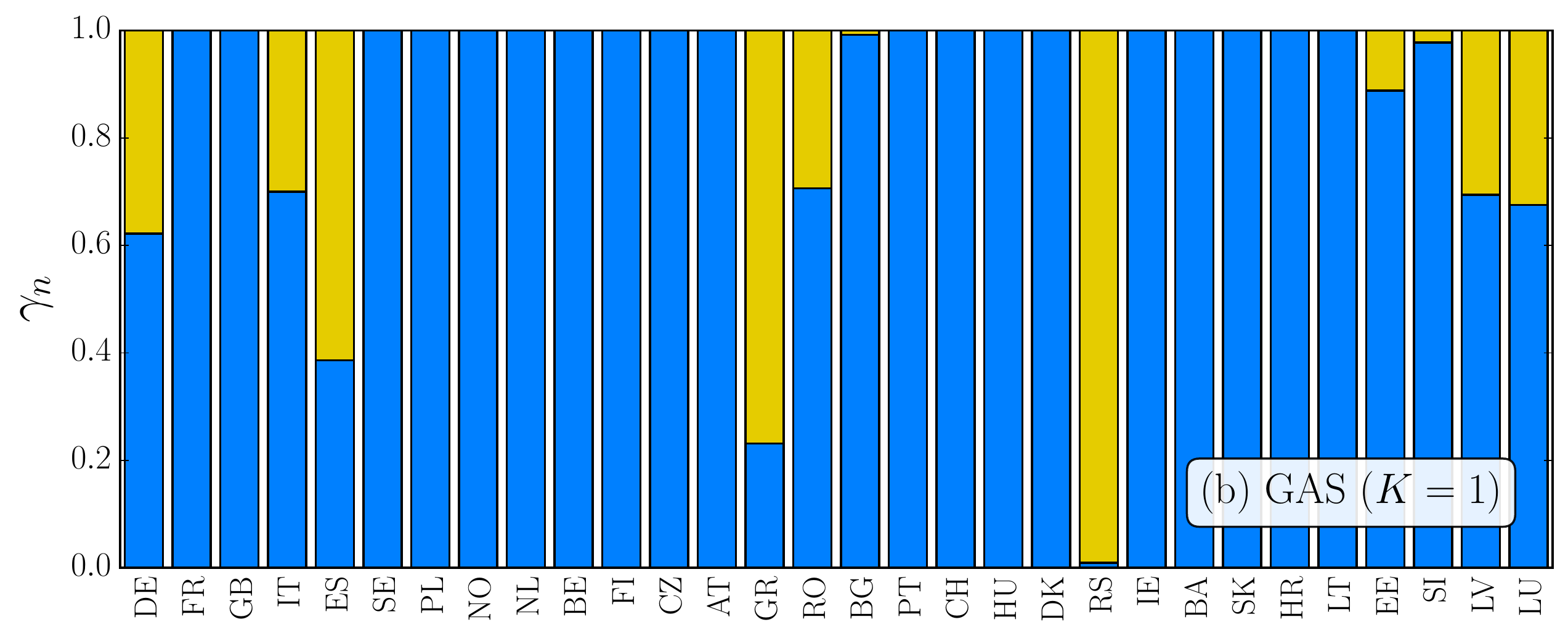}
    %\caption{GAS layout constrained by K = 1.}
    \vspace{-1em}
  \end{subfigure}
  \begin{subfigure}{\columnwidth}
    \includegraphics[width = \chromowidth, center]{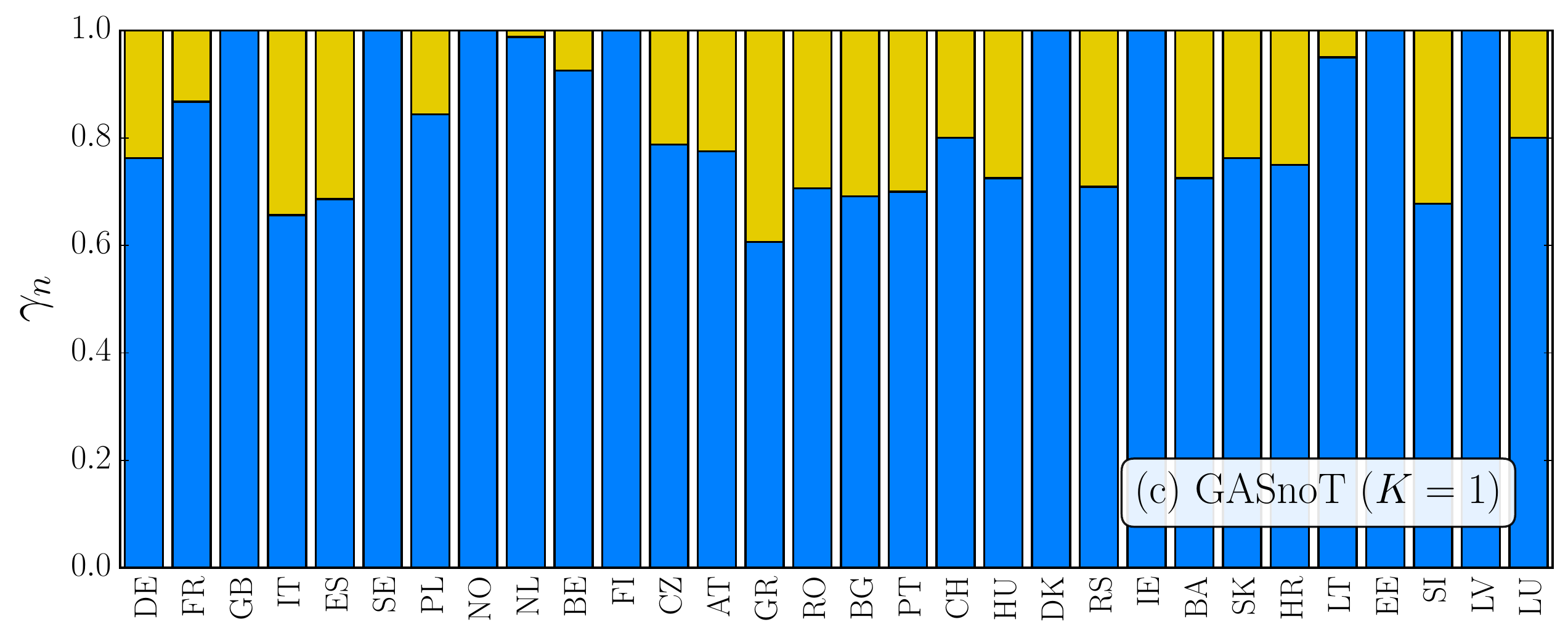}
    %\caption{GASnoT layout constrained by K = 1.}
    \vspace{-1em}
  \end{subfigure}
\caption{
Comparison of (a) the optimal homogeneous layout HOM with the optimised (b) GAS and (c) GASnoT layouts constrained by $K = 1$.
}
\label{fig:K1layouts}
\end{figure}

\begin{figure}[t]  \centering
\includegraphics[width = \columnwidth]{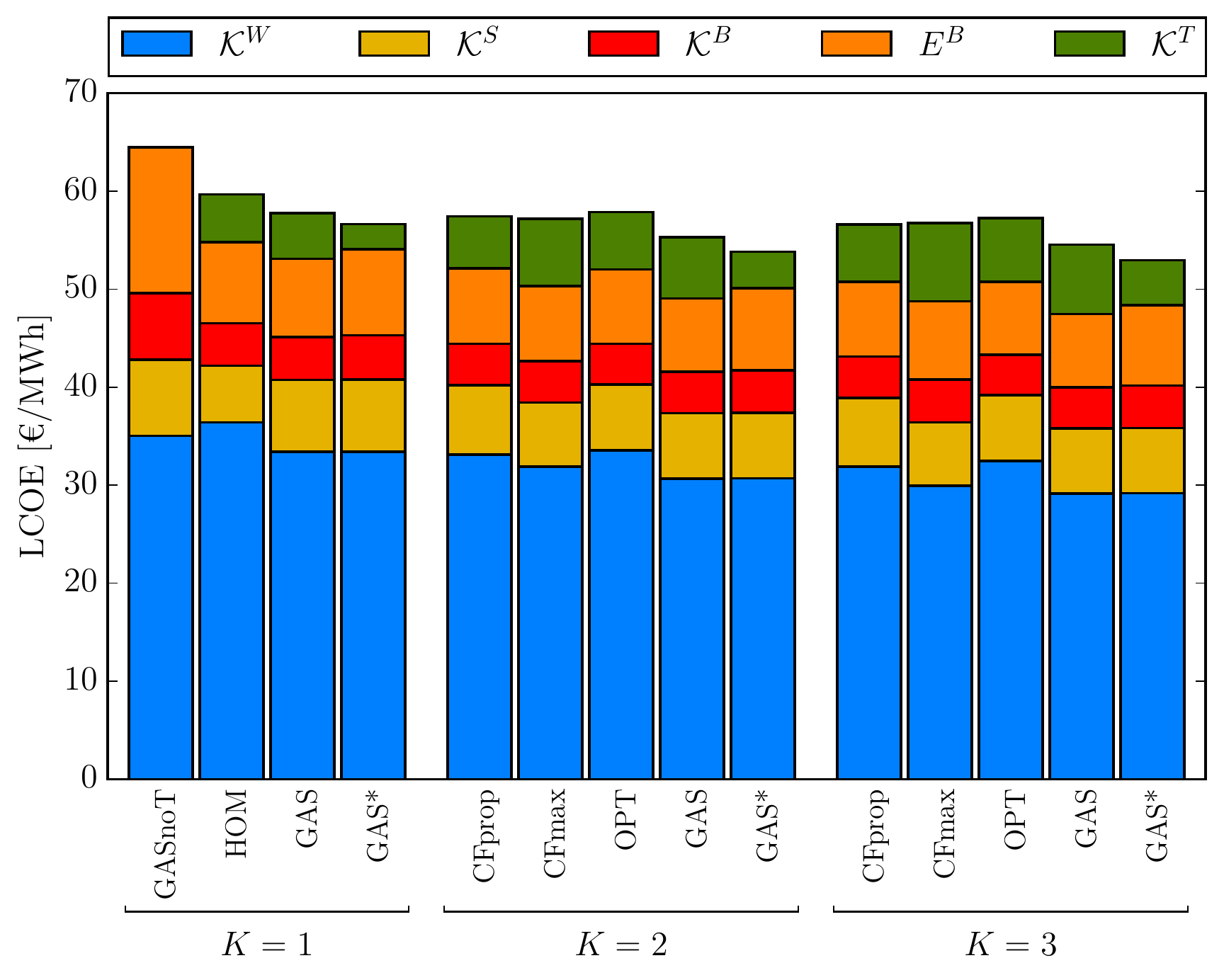}
\vspace*{-0.5em}
\caption{
Componentwise \gls{lcoe} for the optimal CFprop, CFmax, OPT, GAS and GAS* layouts for $K=1$ (left), 2 (middle) and 3 (right). The $K=1$ layout GASnoT without transmission is shown as reference.
}  \label{fig:cost}
\end{figure}

The homogeneous layout producing the minimum LCOE at $\alpha_{EU} = 0.90$ is denoted as the 'HOM' layout. It is illustrated in Figure \ref{fig:K1layouts}a. Its total LCOE amounts to 59.7 \euro/MWh. The componentwise LCOE corresponding to the wind, solar, backup and transmission parts are listed in the third column of Table \ref{tab:cost} and graphed as the second bar in Figure \ref{fig:cost}. Wind power dominates the overall LCOE. Its contribution amounts to 61\%, and is followed by 21\% from backup, 10\% from solar and 8\% from transmission.

\begin{table*}[t!]  \centering
\caption{
Componentwise \gls{lcoe} for the optimal CFprop, optimal CFmax, optimal OPT, optimised GAS and optimised GAS* layouts for $K=1$ (left), 2 (middle) and 3 (right). Note that the $K=1$ layouts CFprop, CFmax and OPT are identical and denoted as HOM. The $K=1$ layout GASnoT without transmission is listed as reference. All costs are given in \euro/MWh.
}
\label{tab:cost}
\resizebox{\textwidth}{!}{%
\begin{tabular}{l|r|rrr|rrrrr|rrrrr}
    \toprule
    & \multicolumn{4}{c|}{$K=1$} & \multicolumn{5}{c|}{$K=2$} & \multicolumn{5}{c}{$K=3$} \\
    \midrule
    & $\mathbf{GASnoT}$
    & $\mathbf{HOM}$    & $\mathbf{GAS}$     & $\mathbf{GAS^*}$
    & $\mathbf{CFprop}$  & $\mathbf{CFmax}$ & $\mathbf{OPT}$ & $\mathbf{GAS}$ & $\mathbf{GAS^*}$
    & $\mathbf{CFprop}$  & $\mathbf{CFmax}$ & $\mathbf{OPT}$ & $\mathbf{GAS}$ & $\mathbf{GAS^*}$ \\
    \midrule
    $\alpha_{EU}$                    &0.86 &0.90 &0.84 &0.84 &0.86 &0.87 &0.87 &0.83 &0.83 &0.85 &0.86 &0.87 &0.82 &0.82 \\
    \midrule
    LCOE($\mathcal{K}^{W}$) &35.0 &36.4 &33.4 &33.4 &33.1 &31.9 &33.6 &30.7 &30.7 &31.9 &30.0 &32.5 &29.1 &29.2 \\ % Onshore wind
    LCOE($\mathcal{K}^{S}$)  &7.8   &5.8   &7.3   &7.4   &7.1   &6.6   &6.7   &6.7   &6.7   &7.0   &6.5   &6.7   &6.7   &6.7   \\ % Solar
    LCOE($\mathcal{K}^{B}$)  &6.8   &4.3   &4.4   &4.5   &4.2   &4.2   &4.2   &4.2   &4.3   &4.2   &4.4   &4.1   &4.2   &4.3   \\ % Backup
    LCOE($E^{B}$)                  &14.9 &8.3   &8.0   &8.8   &7.7   &7.7   &7.6   &7.5   &8.4   &7.6   &8.0   &7.4   &7.5   &8.2   \\ % Fuel
    LCOE($\mathcal{K}^{T}$)  &0.0   &4.9   &4.7   &2.6   &5.3   &6.8   &5.9   &6.2   &3.7   &5.9   &8.0   &6.5   &7.1   &4.6   \\ % Transmission
    \midrule
    LCOE(total)                        &64.5 &59.7 &57.8 &56.6 &57.4 &57.2 &57.9 &55.3 &53.8 &56.6 &56.8 &57.4 &54.5 &53.0  \\
    \bottomrule
\end{tabular}}
\end{table*}

Contrary to the HOM layout, the $K=1$ GAS layout is no longer strictly homogeneous. Of course, all renewable penetrations are still equal to $\gamma_n=1$, but as a result of the optimisation the wind-solar mixing parameters become heterogeneous. This is illustrated in Figure \ref{fig:K1layouts}b. Two-thirds of the countries are wind-only with $\alpha_n=1$. The remaining countries have a significant share of solar. For some of those this was to be expected. Spain, Greece, Italy, Romania and Serbia have very large solar capacity factors. See again Table \ref{tab:capacity-factors}. However, other solar-rich countries, like Portugal, Bulgaria, Bosnia and Croatia, are not amongst them. Instead, Germany is also assigned a significant share of solar, although its solar capacity factor is only average. By taking a closer inspection of Table \ref{tab:capacity-factors} we discover the following empirical finding for the $K=1$ GAS layout: all countries with $\alpha_n=1$ come with a ratio between their solar and wind capacity factor which is smaller than $\mathrm{CF}_n^S / \mathrm{CF}_n^W < 0.65$. The countries with $\alpha_n<1$ have a larger ratio $\mathrm{CF}_n^S / \mathrm{CF}_n^W \geq 0.65$, except for the three smallest countries Estonia, Latvia and Luxembourg.

Compared to the HOM layout, the $\alpha$-heterogeneity of the $K=1$ GAS layout is able to reduce the total LCOE by 3\%. This is mostly a consequence of the reduced combined component costs for wind and solar power. Note, that the overall mixing parameter $\alpha_{EU} = \sum_n \alpha_n \langle L_n \rangle / \langle L_{EU} \rangle$ has also slightly reduced from 0.90 (HOM) to 0.84 (GAS). See the fourth column of Table \ref{tab:cost} and the third bar of Figure \ref{fig:cost}. The costs for backup and transmission have not changed much; \highlight{which is also apparent from the rightmost panel of} Figure \ref{fig:overview}.

All $K=1$ layouts discussed so far include the transmission infrastructure. It is also interesting to compare them to an optimised layout without transmission. No exports and imports would then be possible and the injection pattern $P_n(t)$ would always be zero. No transmission investment would be needed and the respective componentwise LCOE would be zero. However, the countries then have to balance their mismatches all by themselves, and this in turn requires more backup infrastructure with higher respective componentwise LCOE. For the GAS layout without the transmission infrastructure, which for clarity we denote as GASnoT, the total LCOE turns out to be 64.5 \euro/MWh. Compared to the HOM layout, the combined LCOE components for wind and solar power generation are almost the same, but the increase of the LCOE components for the backup power generation and capacity is significantly larger than the disappearance of the transmission component. See again Figure \ref{fig:overview}, Table \ref{tab:cost} and Figure \ref{fig:cost}. The total LCOE of the GASnoT layout is 8\% and 11.5\% larger than for the HOM and GAS layout respectively. This clearly demonstrates the benefit of transmission \cite{Rodriguez2014,Sensitivity}.

The GAS and GASnoT layouts are obtained from two independent optimisation efforts. This explains why the two layouts are actually quite different in the distribution of the wind and solar resources. Figure \ref{fig:K1layouts}c illustrates the resulting wind-solar mixing parameters for the GASnoT layout. Contrary to the more extreme GAS layout, the majority of the countries comes with a mix below $\alpha_n=1$ and well above $0$. Only the most northern countries turn out to be wind-only. However, on average the mixing parameter $\alpha_{EU}=0.86$ for the GASnoT layout is again close to $\alpha_{EU}=0.84$ for the GAS layout.

%*****************************
\subsection{$K=2$ layouts}
\label{subsec:ResultsK2}

\begin{figure}[t]  \centering
  \begin{subfigure}{\columnwidth}
    \includegraphics[width = 0.80 \columnwidth, center]{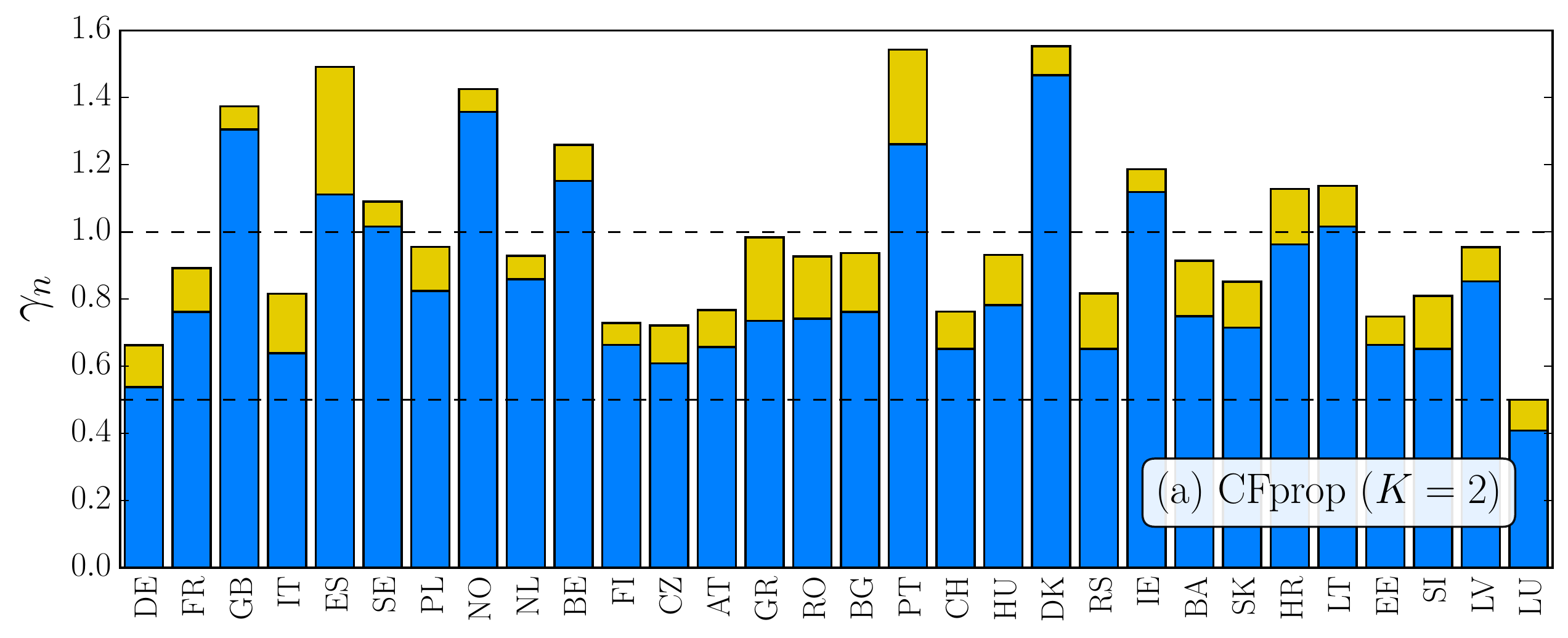}
    %\caption{$\beta$ layout constrained by $K = 2$.}
    \vspace{-1em}
  \end{subfigure}
  \begin{subfigure}{\columnwidth}
    \includegraphics[width = 0.80 \columnwidth, center]{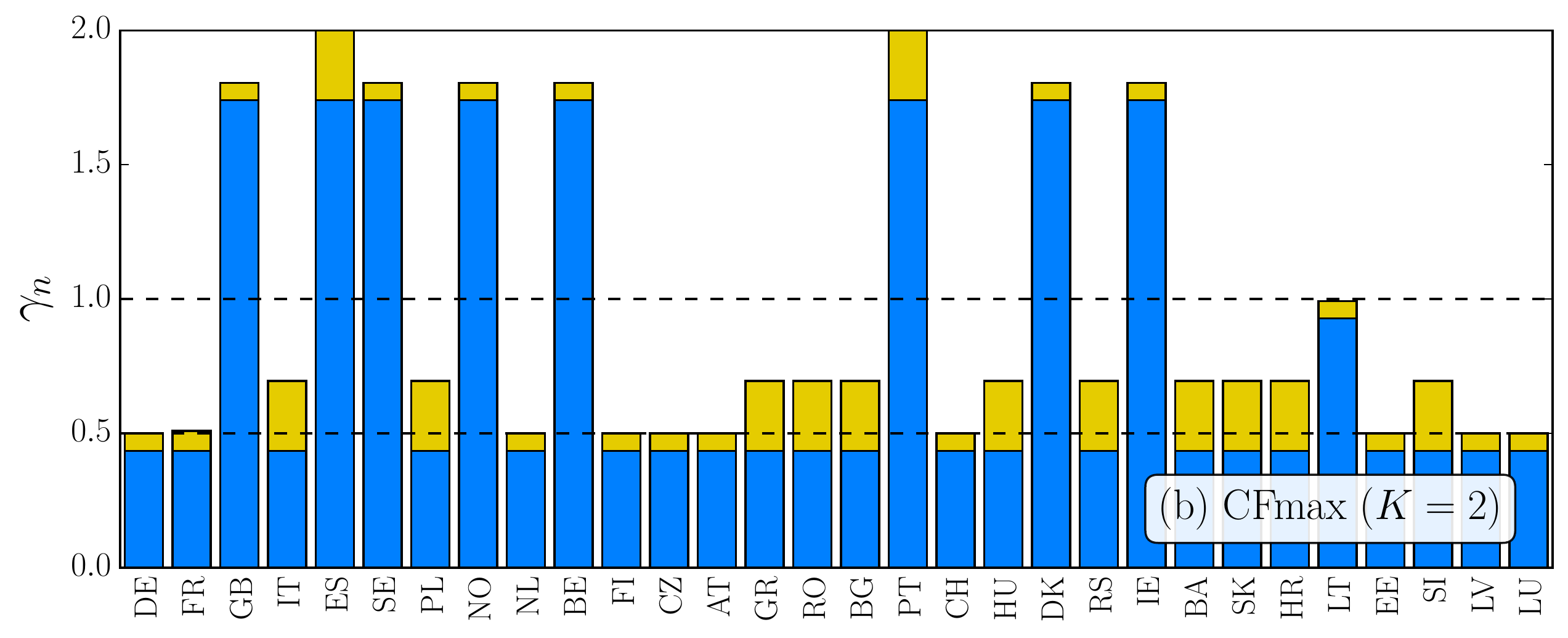}
    %\caption{CF layout constrained by $K = 2$.}
    \vspace{-1em}
  \end{subfigure}
  \begin{subfigure}{\columnwidth}
    \includegraphics[width = 0.80 \columnwidth, center]{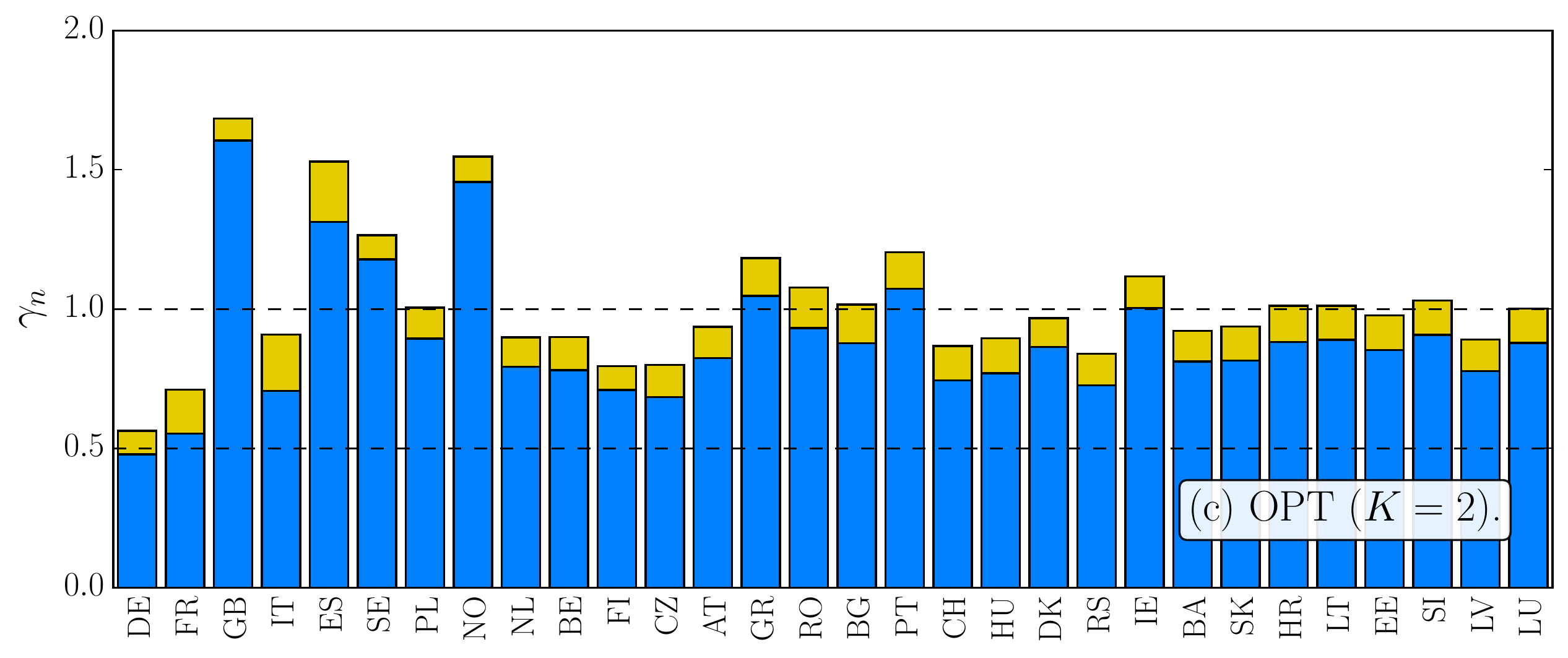}
    %\caption{OPT layout constrained by $K = 2$.}
    \vspace{-1em}
  \end{subfigure}
  \begin{subfigure}{\columnwidth}
    \includegraphics[width = 0.80 \columnwidth, center]{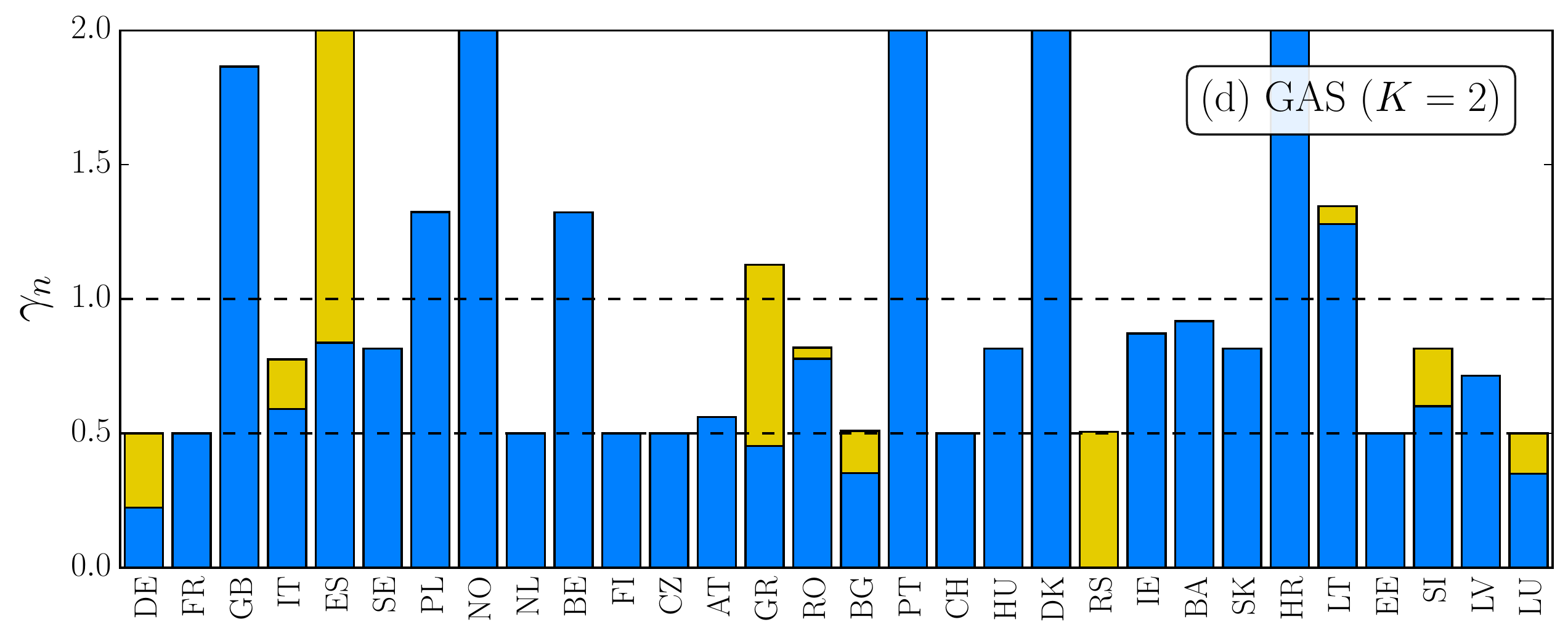}
    %\caption{GAS layout constrained by K = 2.}
    \vspace{-1em}
  \end{subfigure}
\caption{
Comparison of different layouts constrained by $K=2$: (a) CFprop, (b) CFmax, (c) OPT  and (d) GAS.
}  \label{fig:K2layouts}
\end{figure}

More heterogeneity is introduced once $K$ is chosen to be larger than one. Figures \ref{fig:K2layouts}a-c illustrate the optimal heuristic CFprop, CFmax and OPT layouts for $K=2$. Their respective $\alpha_{EU}$ values are 0.86 -- 0.87 (see Table \ref{tab:cost}), and have been fixed by minimising the LCOE (see Figure \ref{fig:overview}d). The general $\alpha_{EU}$-dependence of the other infrastructure measures are illustrated in Figure \ref{fig:overview}a-c. The backup energies required for the three layouts are quasi identical, and no difference is seen to the $K=1$ HOM layout. Also the backup capacities are almost identical for the three layouts, and are slightly less than for the $K=1$ HOM layout. Differences are observed for the transmission capacities. The CFprop layout comes with the smallest transmission capacities, followed by the OPT layout. The CFmax layout has the largest transmission capacities because its heterogeneity is the largest. All $K=2$ layouts are found to have larger transmission capacities than the respective $K=1$ layouts.

The total LCOE of the three heuristic $K=2$ layouts are inbetween 57.2 -- 57.9 \euro/MWh. See columns 6-8 in Table \ref{tab:cost} and bars 5-7 in Figure \ref{fig:cost}. This is very close to the value 57.8 \euro/MWh found for the $K=1$ GAS layout. In this respect, the larger heterogeneity of the $K=2$ layouts do not represent a clear cost advantage when compared to the $K=1$ GAS layout, which is homogeneous in the renewable penetration parameters $\gamma_n$. The situation changes once the optimised $K=2$ GAS layout is considered, which is exemplified in Figure \ref{fig:K2layouts}d. It exploits the wind resources over Europe in a more efficient way and reduces the wind component in the LCOE; consult column 9 of Table \ref{tab:cost} and bar 8 in Figure \ref{fig:cost}. This reduces the total LCOE to 55.3 \euro/MWh.

\begin{figure}[t]  \centering
\includegraphics[width = \columnwidth]{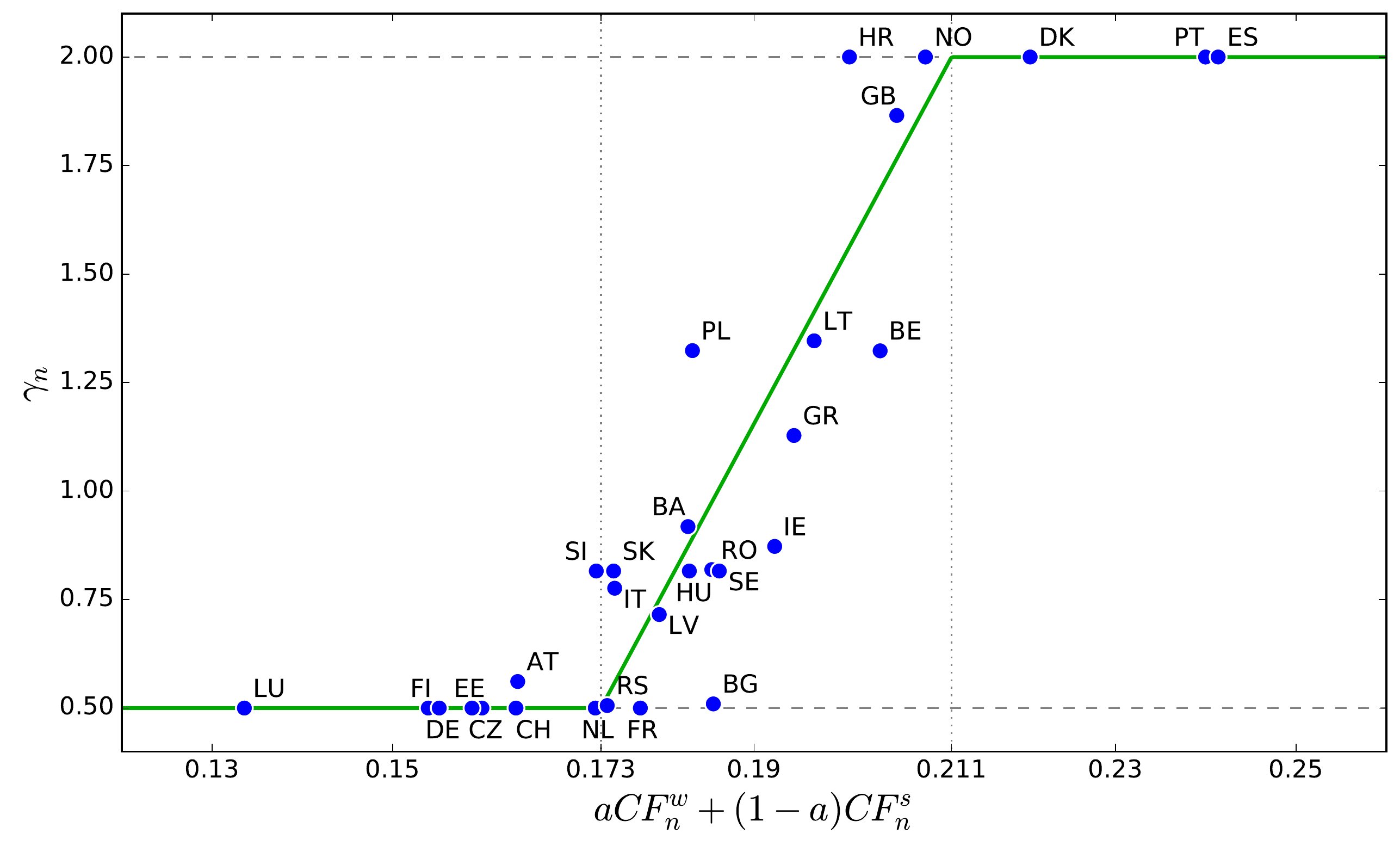}
\vspace*{-0.5em}
\caption{
Renewable penetration parameters $\gamma_n$ from the $K=2$ GAS layout as a function of the effective capacity factor $CF_n^{eff}$ defined in Equation (\ref{eq:CFeff}). The continuous and piecewise linear green function represents the heuristic law (\ref{eq:GammaRegression}) with least-square-fitted parameters $a=0.596$, $CF_1=0.173$ and $CF_2=0.211$.
}  \label{fig:regression}
\end{figure}

The overall renewable penetration of the $K=2$ GAS layout is $\gamma_{EU}=1$; consult again Equation (\ref{eq:norm}). However, the individual renewable penetration parameters now scatter within $0.5 \leq \gamma_n \leq 2$. As can be seen in Figure \ref{fig:K2layouts}d, their distribution is extremely heterogenous. For half of the countries they are either $\gamma_n=2$ or $\gamma_n=0.5$, and for the other countries just somewhere in-between. A more carefull inspection reveals an approximate heuristic law, which expresses the renewable penetration parameters
% -------------------
\begin{equation}  \label{eq:GammaRegression}
   \gamma_n
      =   \left\{ \begin{array}{ll}
      		1/K
		&  (CF_n^\textrm{eff} \leq CF_1)  \\
		(K-\frac{1}{K}) \frac{CF_n^\textrm{eff}-CF_1}{CF_2-CF_1} + \frac{1}{K}
		&  (CF_1 \leq CF_n^\textrm{eff} \leq CF_2)  \\
		K
		&  (CF_n^\textrm{eff} \geq CF_2)
           \end{array} \right.
\end{equation}
% -------------------
as a continuous and piece-wise linear function of an effective capacity factor
% -------------------
\begin{equation}  \label{eq:CFeff}
   CF_n^\textrm{eff}
      =   a CF_n^W + (1-a) CF_n^S  \; .
\end{equation}
% -------------------
A least-square fit is shown in Figure \ref{fig:regression}.

The overall mixing parameter $\alpha_{EU}=0.83$ of the $K=2$ GAS layout is almost the same as for the $K=1$ GAS layout. Both layouts also have in common that 20 out of the 30 countries come with $\alpha_n=1$. The five largest of the $\alpha_n<1$ countries with a non-zero solar component are also identical.

It is worth to take again a quick look at Figure \ref{fig:portfolios}. It shows that for $K=2$ the optimal CFprop, CFmax and OPT layouts have more or less the same close-to-minimum standard deviation of the overall mismatch (\ref{eq:DeltaEU}) as the optimised GAS layout. This indicates that a minimised mismatch standard deviation serves as a good measure to determine an optimal infrastructure \cite{HOL08}. However, it is still a rough measure, since it does not allow to finetune the minimum-cost infrastructure.

%*****************************
\subsection{$K=3$ layouts}
\label{subsec:ResultsK3}

\begin{figure}[t]  \centering
  \begin{subfigure}{\columnwidth}
    \includegraphics[width = \chromowidth, center]{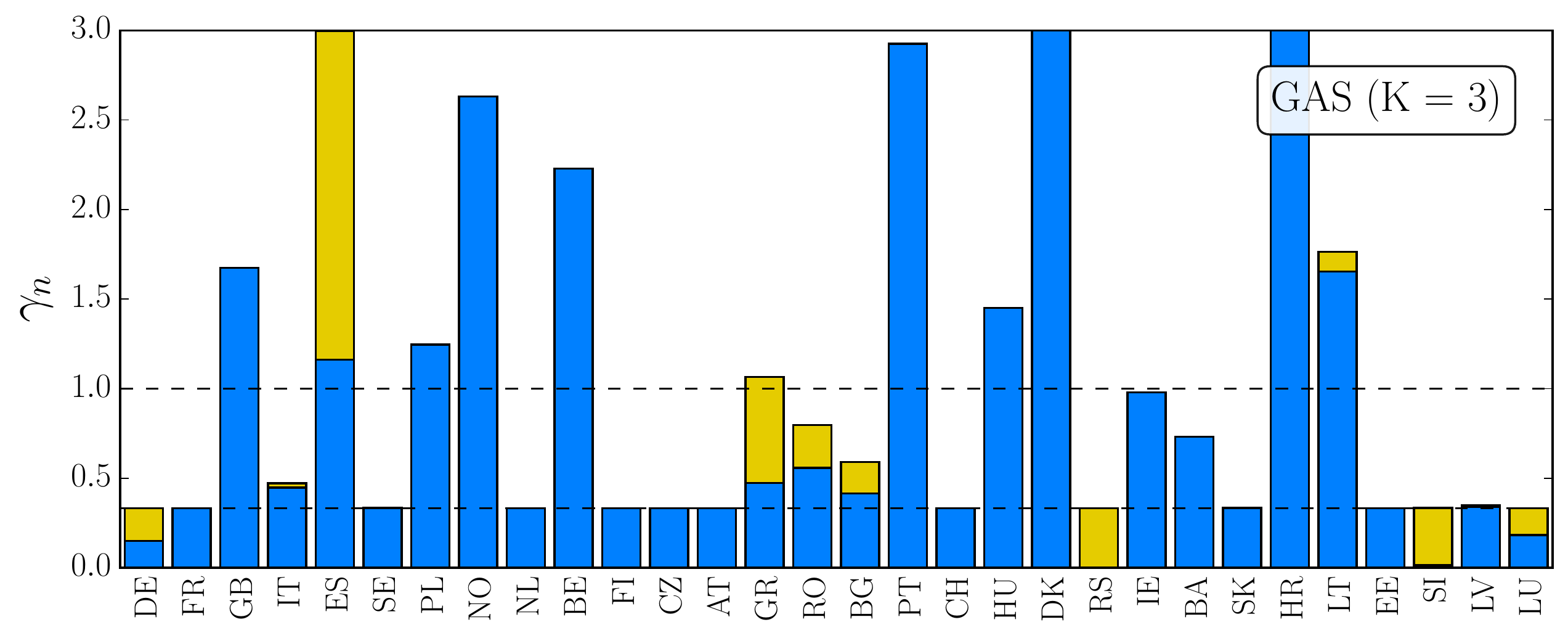}
    \vspace{-1em}
  \end{subfigure}
\caption{
GAS layout constrained by $K=3$.
}  \label{fig:K3layouts}
\end{figure}

For $K=3$ the GAS algorithm has more freedom to optimise the heterogeneous layout and to reduce the overall LCOE\highlight{, see \eqref{eq:k-factor}}. The resulting layout is depicted in Figure \ref{fig:K3layouts}. It has some similarity to the $K=2$ GAS layout, but of course the $K=3$ GAS layout is even more extreme. Its overall wind-solar mixing parameter $\alpha_{EU}=0.82$ is almost the same as for the $K=2$ counterpart. The overall cost reduction turns out to be small. As can be seen in Table \ref{tab:cost}, the total LCOE for the $K=2$ and $K=3$ GAS layouts are 55.3 and 54.5\euro/MWh, respectively. This small cost reduction is mainly caused by the opportunity to allocate more wind resources to the sites with a very high capacity factor, and it is weakened to some extend by slightly increased costs for the transmission component; compare column 14 with column  9 in Table \ref{tab:cost}.

Bulk results for the optimal heuristic $K=3$ layouts CFprop, CFmax and OPT are also listed in Table \ref{tab:cost} and Figure \ref{fig:cost}. Their layouts are also found to be wind-dominated, with nearly the same $\alpha_{EU}$ values as for the respective GAS layout. The LCOE for these three heuristic layouts are larger than for the $K=3$  GAS layout. This of course was to be expected. However, their LCOE also turn out to be slightly larger than for the less heterogeneous $K=2$ GAS layout.

Another reason that the GAS optimisation might have been better than the heuristic layouts is that the GAS algorithm sees not just the capacity factors at each site, like the heuristic layouts, but also the geographical variation of the temporal generation pattern, which the GAS algorithm can exploit to shape the \gls{vres} generation pattern towards the load. However if this was the reason, the backup generation costs would have decreased from the heuristic to the GAS layout, which they do not. This suggests that the GAS optimisation's success really lies with the free exploitation of capacity factors.

\subsection{Transmission capacities}
\label{subsec:TransCap}

So far, only the total contribution of the transmission capacities to the overall LCOE have been discussed for various system layouts in Table \ref{tab:cost} and Figure \ref{fig:cost}. Its geographic distribution has not yet been specified. This will be done in this Subsection, but not right away. At first we will investigate a procedure which further reduces the overall LCOE by reducing the transmission capacities to some extend.

The transmission capacities defined in Equation (\ref{eq:link-cap}) have been derived from unconstrained power flows. They are determined by the most extreme flow events, which typically occur between countries with a large energy deficit and others with a large excess. These events are not expected to overlap with other extreme events when all countries face a large energy deficit. The latter determine the required backup capacities. Consequently, it can be expected that a modest reduction of the total transmission capacities will not, or at least not much, affect the total backup capacities and the total backup energy, and will lower the overall LCOE.

The synchronised balancing scheme (\ref{eq:synch}) presented in Section \ref{subsec:network} is based on unconstrained power flows. In order to include constrained power flows, a generalisation is needed:
% -------------------
\begin{equation}  \label{eq:step1}
  \begin{aligned}
    & \underset{\mathbf{B}}{\text{min}}
    & & \sum_n \frac{\paren{B_{n}(t)}^{2}}{\mean{L_{n}}} \\
    & \text{s.t.}
    & & \sum_{n} P_{n} (t) = 0 \\ %F_{l}^{-} \leq F_{l} \leq F_{l}^{+}
    & \text{s.t.}
    & & -\mathcal{K}_l^{conT} \leq F_{l} (t) = \sum_n H_{ln} P_n(t) \leq \mathcal{K}_l^{conT}  \; .
  \end{aligned}
\end{equation}
% -------------------
The objective is to minimise the expression in the first line, taking into account the two constraints of the second and third line.
$\mathcal{K}_{l}^{conT}$ denotes the constrained transmission capacity of line $l$. In the limit of unconstrained flows, where the second constraint can be discarded, the objective (\ref{eq:step1}) can be rewritten as $\underset{\mathbf{B}}{\text{min}} \sum_n (B_n^2(t)/\mean{L_n} - \lambda P_n)$ with the method of Lagrange multipliers and leads to the solution (\ref{eq:synch}). For the following we will downscale the unconstrained transmission capacities from (\ref{eq:link-cap}) by a uniform scaling parameter $\zeta$ to obtain the constrained transmission capacities
% -------------------
\begin{equation}  \label{eq:Tcon}
   \mathcal{K}_l^{conT} =  \zeta \mathcal{K}_l^{T}  \; .
\end{equation}
% -------------------

\begin{figure}[t]  \centering
\includegraphics[width = \columnwidth]{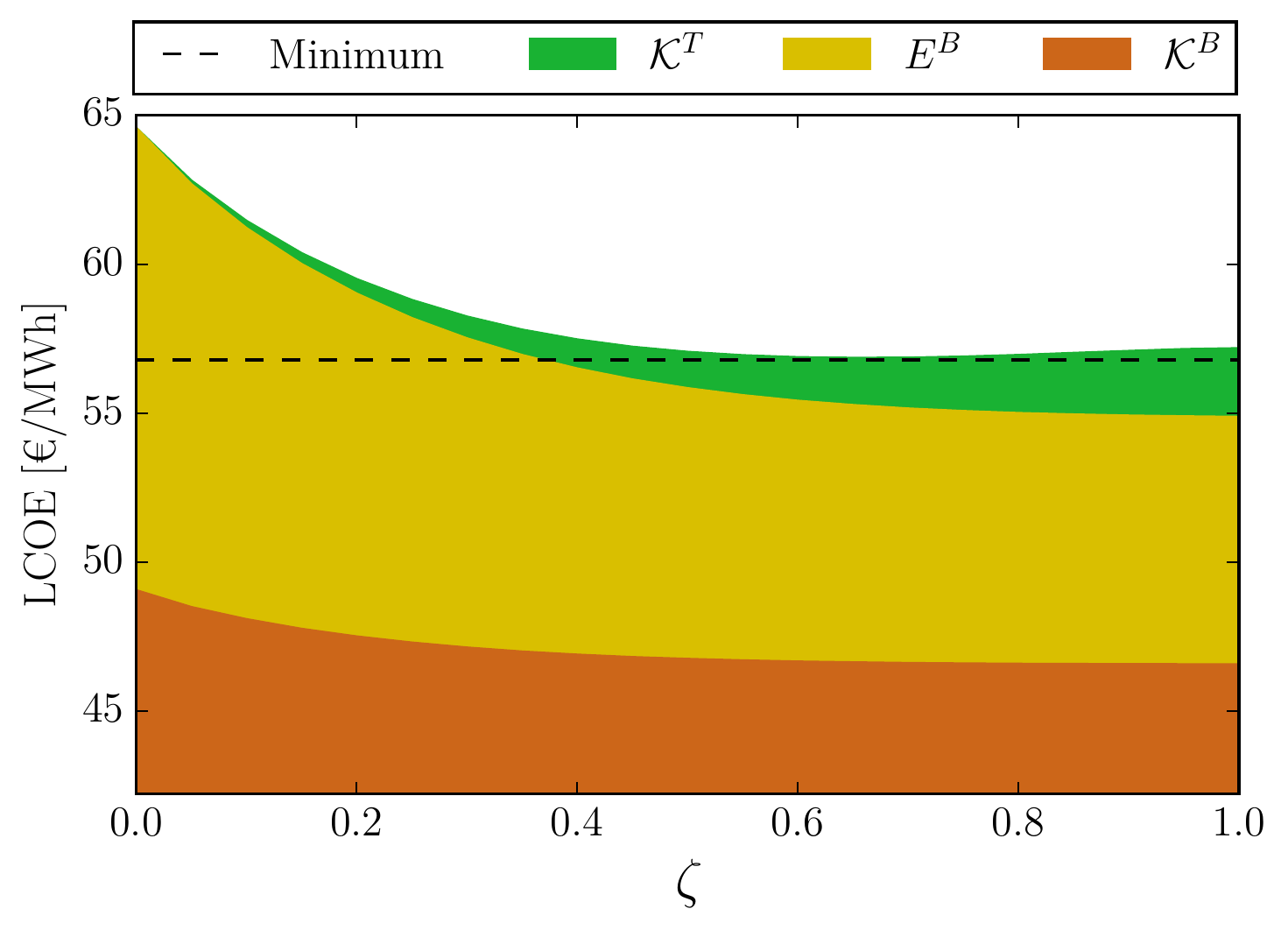}%
\vspace*{-0.5em}%
\caption{
Non-\gls{vres} components of the \gls{lcoe} as a function of the scaling parameter $\zeta$. The dashed line indicates the minimum leading to the lowest \gls{lcoe}. The calculations were performed using the $K=2$ GAS layout at $\zeta=1$. The \gls{vres} part, which does not depend on $\zeta$, is not shown; it consists of 30.7\euro/MWh for wind and 6.7\euro/MWh for solar.
}  \label{fig:transmission-lcoe}
\end{figure}

Figure \ref{fig:transmission-lcoe} illustrates the dependence of the LCOE on the transmission constraints by taking the unconstrained transmission capacities of the $K=2$ GAS layout and scaling them down by the uniform factor $\zeta$. At first, as $\zeta$ decreases, the LCOE also decreases. A minimum is found at $\zeta=0.60$. For the $K=1$ and $K=3$ GAS layouts the minimum is found at the optimal values $\zeta=0.55$ and $0.65$, respectively. If the transmission capacities are downscaled further the LCOE starts to increase again due to increasing requirements for backup energy and backup capacity.

Table \ref{tab:cost} lists also the modified GAS layouts resulting from the optimal scaling parameters. For clarity, we denote them as GAS$^*$ layouts. Compared to the GAS layouts, the transmission contribution to the total LCOE is reduced and the backup contributions are slightly increased. The wind and solar components of the GAS and GAS$^*$ layouts are of course identical. Compared to the $K=1$ GAS layout, the total LCOE of the $K=1$ GAS$^*$ layout is reduced by 1.2\euro/MWh in absolute units and by 2.1\% in relative units. For $K=2$ and $K=3$ the reductions are 2.7\% and 2.8\%, respectively. The reductions are also illustrated in Figure \ref{fig:cost}.

\begin{figure}[t]  \centering
\includegraphics[width = \columnwidth]{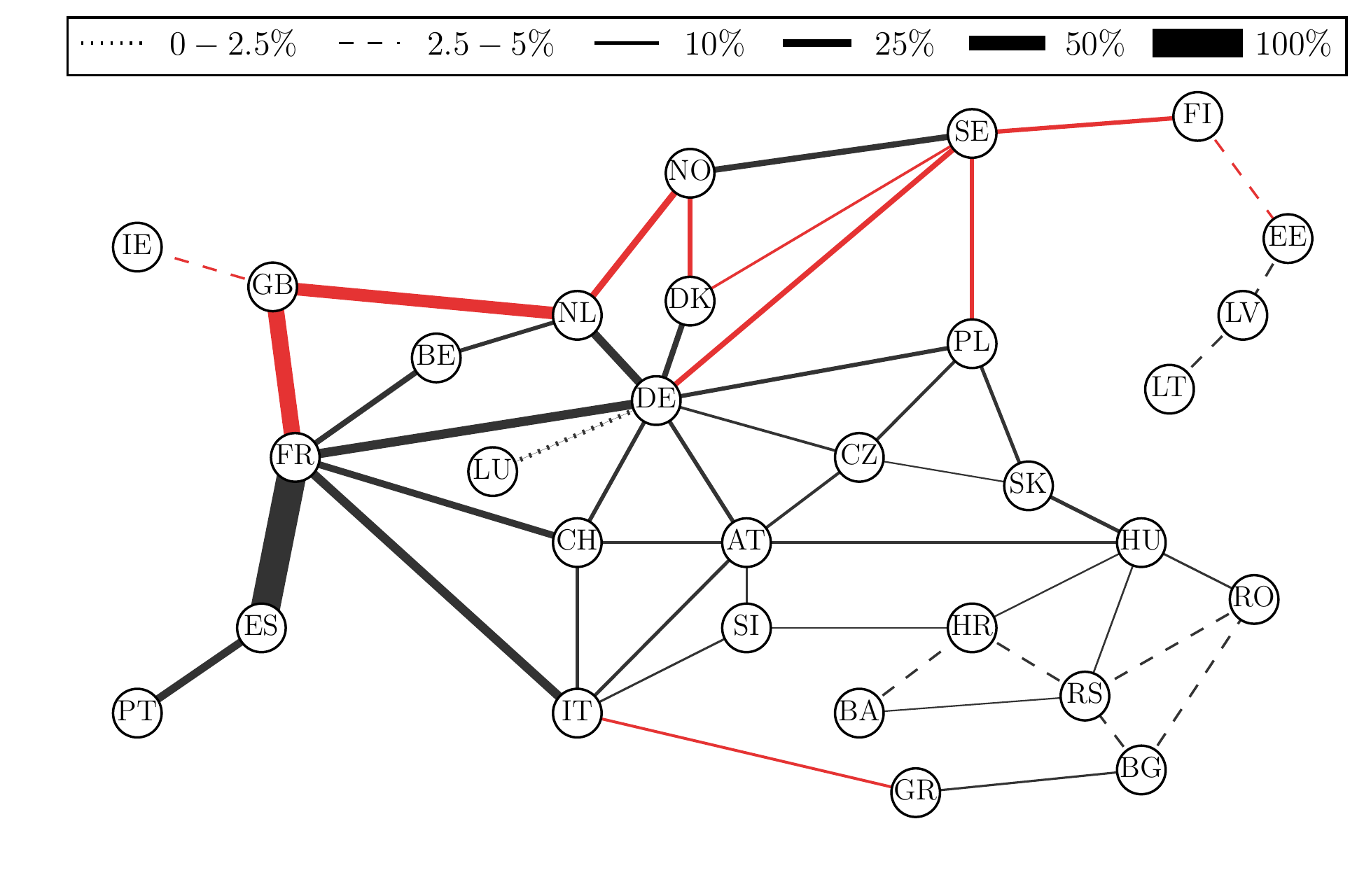}
\caption{
Geographic distribution of the transmission capacities for the $K=2$ GAS$^*$ layout. AC links are shown in black while HVDC links are shown in red. Link capacities are indicated relative to the highest capacity, which is 68 GW between France and Spain.
}  \label{fig:links}
\end{figure}

The geographic distribution of the transmission capacities for the $K=2$ GAS$^*$ layout is shown in Figure \ref{fig:links}. The transmission capacities are not homogeneously distributed across the network. By far the strongest links are attached to Spain and Great Britain, which are the two largest countries with severe renewable excess generation. Links to their second neighbours with big deficits in renewable power generation, in particular Germany and Italy, also turn out to be quite strong. The more expensive HVDC transmission lines are utilised less extensively.

\section{Sensitivity analysis}
\label{sec:sensitivity-analysis}

\subsection{Reduced solar cost}
\label{sec:reduced-solar-cost}

\begin{figure*}[t!]  \centering
\includegraphics[trim={0.3cm 0.2cm 0.25cm 0.2cm},clip,width = \textwidth, center]{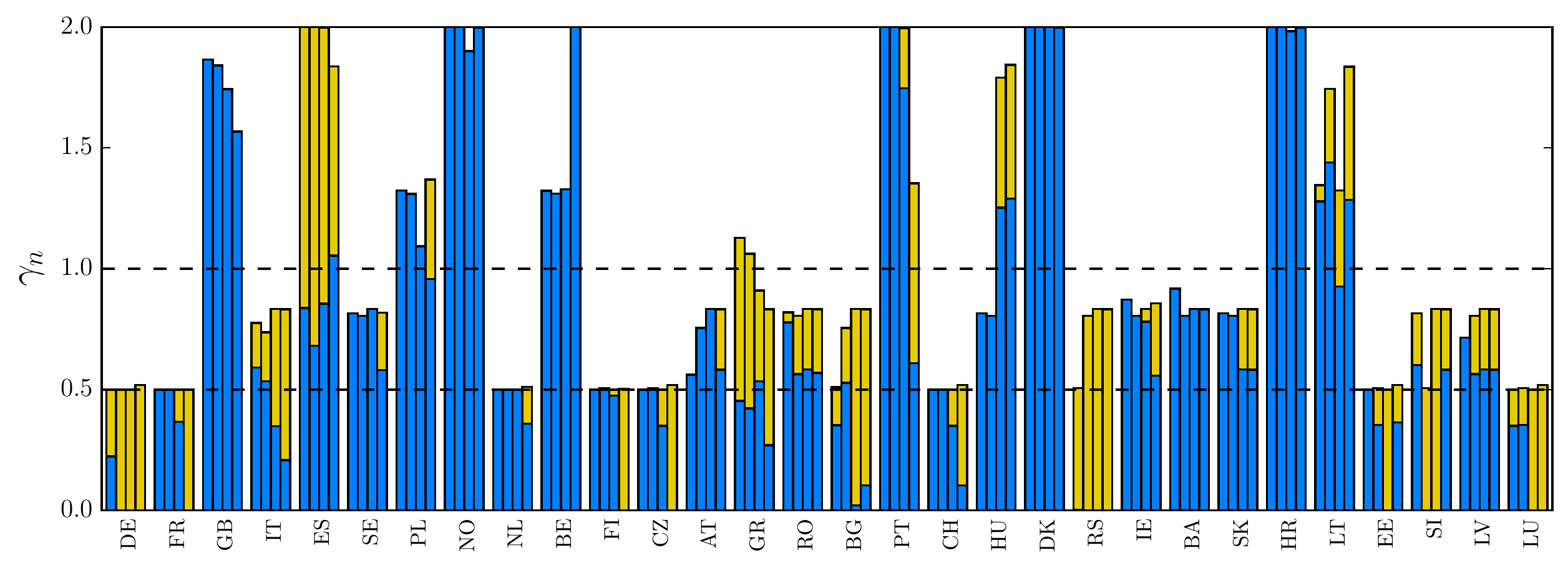}
\caption{
GAS optimised layouts constrained by K = 2 for a solar cost reduction of 0\%, 25\%, 50\% and 75\%, from left to right.
}  \label{fig:layout-solar}
\end{figure*}

For the optimised GAS layouts as well as for the heuristic CFprop, CFmax and OPT layouts the optimal mixing parameter $\alpha_{EU}$ minimising the overall costs is located in the wind dominated region. This is a consequence of the substantially higher costs of solar generation compared to wind. The future price development of solar photovoltaic systems is rather uncertain. To analyse the sensitivity to future price drops in solar cost, we calculate optimised layouts for solar cost reductions of 25\%, 50\% and 75\%. Cost reductions could come from improved production processes, or alternatively from increasing capacity factors. Based on data from \cite{AND15}, the capacity factor can be increased by up to 40\% by applying dual axis tracking compared to the fixed position installation assumed in Table \ref{tab:capacity-factors}, which may offset the higher capital costs of such systems. In addition, studies on increasing the energy conversion efficiency are still being conducted. \highlight{A recent study suggests a huge decrease in the total system cost of PVs in a far future system \cite{irena2}}.

The resulting $K=2$ GAS portfolios are visualized in Figure \ref{fig:layout-solar}. Not surprisingly we find that a decrease in solar cost leads to a continuously increase in totally installed solar capacity. This increase is not found to be equal at all nodes. The main solar electricity supplier, Spain, initially increases its solar capacity, but for the more extreme price reductions decreases it again. It seems more efficient to shift the production to other sites. Spain is the clear leader in terms of solar generation for large solar costs. However, in the case of 50\% solar cost reductions Germany almost produces equal amounts as Spain. This might not appear to be intuitive from the figure as the renewable penetration of Germany is always smaller than for Spain, but the mean load of Germany is more than twice as large as the one of Spain. In the 75\% scenario Germany passes Spain and becomes the main producer of solar power. In this most extreme scenario almost all countries deploy solar resources.

\begin{figure}[t]  \centering
\includegraphics[trim={0cm 0.4cm 0cm 0cm},clip,width = \columnwidth]{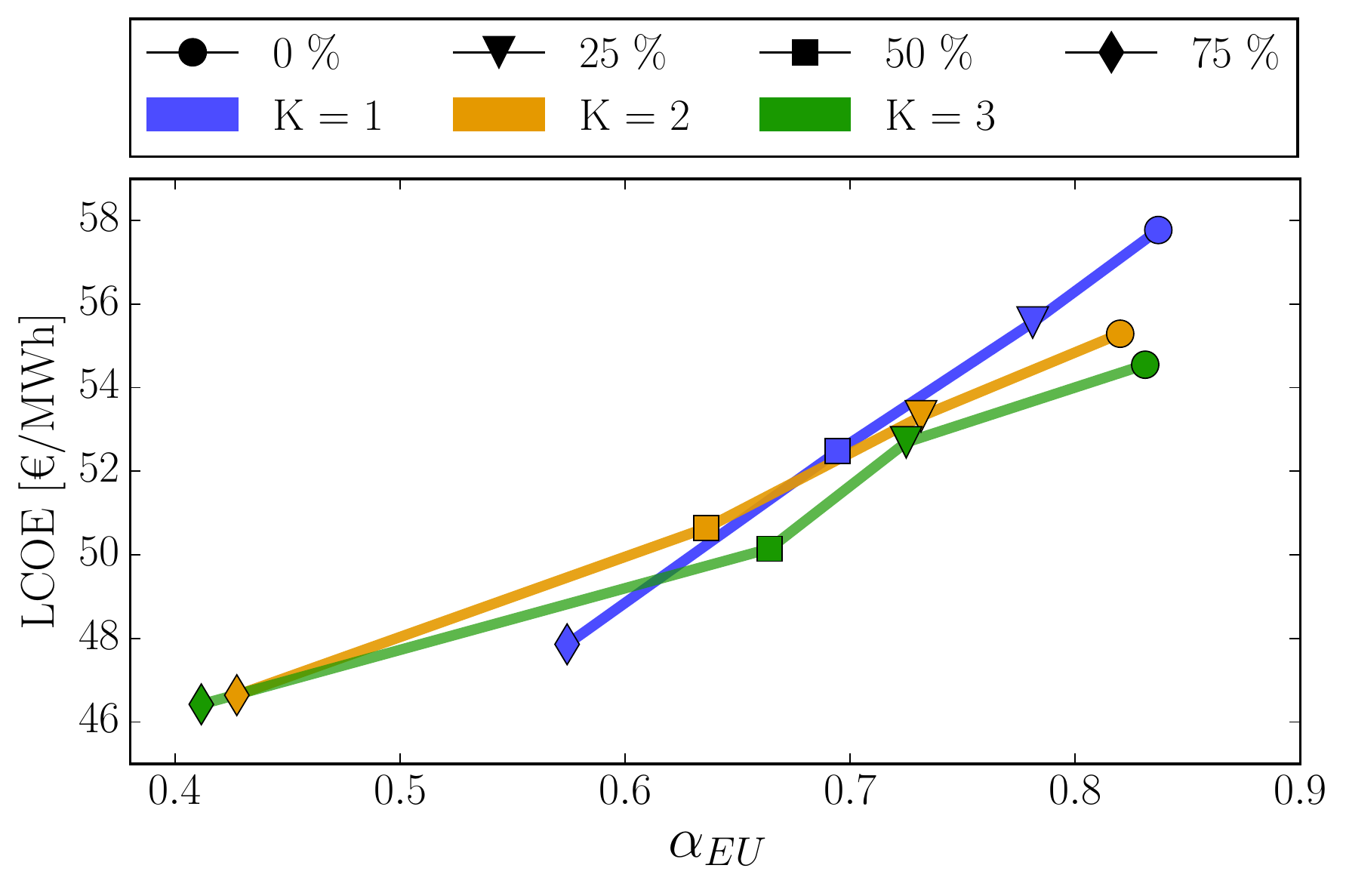}
\caption{
\gls{lcoe} of the GAS optimised layouts when the solar cost is reduced by 25\% (triangle), 50\% (square) and 75\% (diamond). The 0\% scenario (circle) is included as a reference. Different constraints are shown: K = 1 (blue), 2 (yellow) and 3 (green).
}  \label{fig:red-solar}
\end{figure}

We illustrate the change in the associated LCOE due to the cost reductions in the cases of K = 1, 2 and 3 in Figure \ref{fig:red-solar}. For all cases of heterogeneity the associated total European LCOE drops steadily for decreasing solar costs. For a reduction of the solar cost by 25\% the optimal mix is shifted from above $\alpha_{EU}=0.8$ to below this point, and the LCOE values drop by around 2\euro. As the solar cost is reduced by 50\% the optimal mix drops further and lies between 0.6 and 0.7. For $K=2$ the LCOE is reduced by almost 5{\euro} compared to the reference scenario. When reducing the cost of solar by 75\%, solar becomes much cheaper than wind, and the optimal mix is shifted below $\alpha_{EU}=0.5$, indicating a dominant share of solar. Compared to the reference scenario, the LCOE dropped by around 9{\euro} for the case $K=2$. We have to be aware that such large cost reductions for solar photovoltaic systems might not be plausible. A cost reduction is mostly to be expected from material and production costs but not from installation costs.

\subsection{Increased backup cost}
\label{sec:increased-backup-cost}

The future price developments of fossil fuels, which are likely to increase, will affect the cost of electricity. An increase in the cost of gas used by the CCGT generators leads to an increase in the variable operational expenses associated with backup generation. In principle this will also affect the structure of the optimised layouts, but we expect the structural change to be very small. As Figure \ref{fig:overview}a reveals, the mixing parameters $\alpha_{EU}=0.82$-0.84 of the optimised GAS layouts also produce the minimum of the backup energy. Consequently, the structure of the layouts will more or less not change, but of course their LCOE will increase as the gas price increases. This increase is linear. For the $K=2$ GAS layout an increase in backup fuel price to 150\% leads to a LCOE of 59.0\euro/MWh, which is an increase of 6.8\%. An increase to 200\% of the gas price results in a LCOE of 62.8\euro/MWh, which equals an increase of 13.6\%.

The increased backup costs can to some degree be counterbalanced by the sale of curtailment energy. So far we have assumed that curtailed electricity is wasted renewable production. Selling the curtailment energy to other energy sectors like the heating and transportation sector is a promising possibility. The resulting decrease in LCOE depends on the selling price and the amount of electricity sold. Since we are discussing an all-European renewable penetration of $\gamma_{EU}=1$ throughout this paper, the total amount of curtailment energy is identical to the backup energy. Assuming to sell $1/3$ of it at a price of 80\euro/MWh, the LCOE of the $K=2$ GAS layout is reduced to 50.2\euro/MWh, which is a decrease of 9.2\%. Note however, that the sale of curtailment energy might have a slightly bigger impact on the structural change of the optimised GAS layouts than increased backup costs. Since, again, the amount of curtailment energy is equal to the backup energy, Figure \ref{fig:overview}a also illustrates the dependence of the curtailment energy on the mixing parameter $\alpha_{EU}$. For parameter values below $\alpha_{EU}=0.82$-0.84 the curtailment energy increases strongly. Consequently, when taking the sale of curtailment energy into account, a proper layout optimisation will shift to some degree towards smaller mixing parameters.

\subsection{Interpolations towards more and less heterogeneity}
\label{sec:HetInterpolations}

\begin{figure}[t]  \centering
\includegraphics[trim={0cm 0.4cm 0cm 0cm},clip,width = \columnwidth]{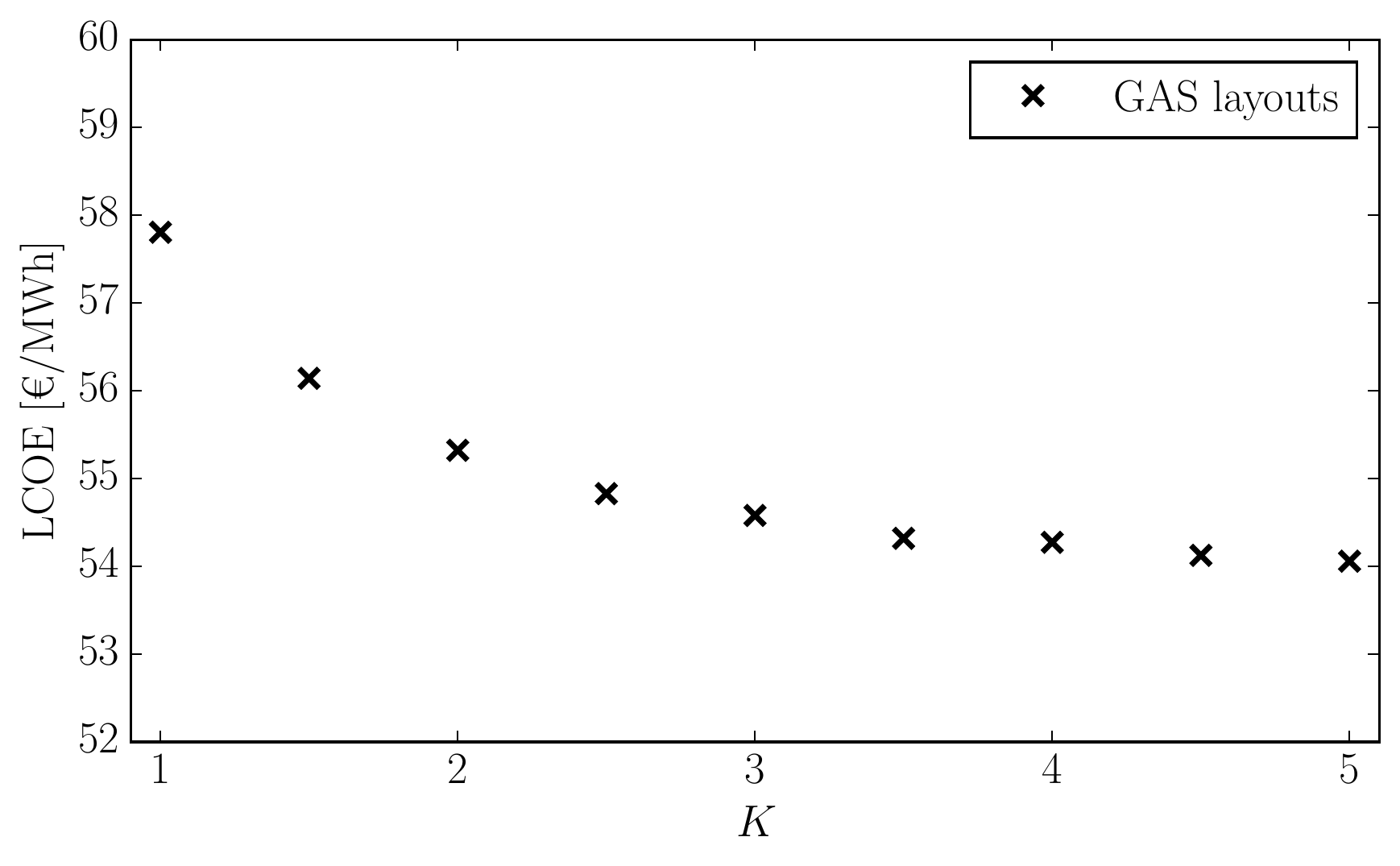}
\vspace*{-0.5em}%
\caption{
LCOE of the optimised GAS layouts as a function of the constraint parameter $1\leq K\leq 5$.
}  \label{fig:GAS1K5}
\end{figure}

As the heterogeneity parameter changes from $K=1$ to 2 and 3, the LCOE of the optimised GAS layouts has decreased further; consult again Table \ref{tab:cost} and Figure \ref{fig:cost}. It is quite natural to ask how much further the LCOE might decrease as $K$ gets even larger. The answer is shown in Figure \ref{fig:GAS1K5}. The LCOE decreases continuously with increasing heterogeneity. However, the benefit of increased heterogeneity becomes smaller and smaller. The increasing cost of transmission leads to a point where it is almost no longer economic beneficial to increase the heterogeneity. The LCOE of 54.5\euro/MWh for the $K = 3$ GAS layout is already very close to the asymptotic value of 54\euro/MWh for very large $K$.

\begin{figure}[t]  \centering
\includegraphics[trim={0cm 0.4cm 0cm 0cm},clip,width = \columnwidth]{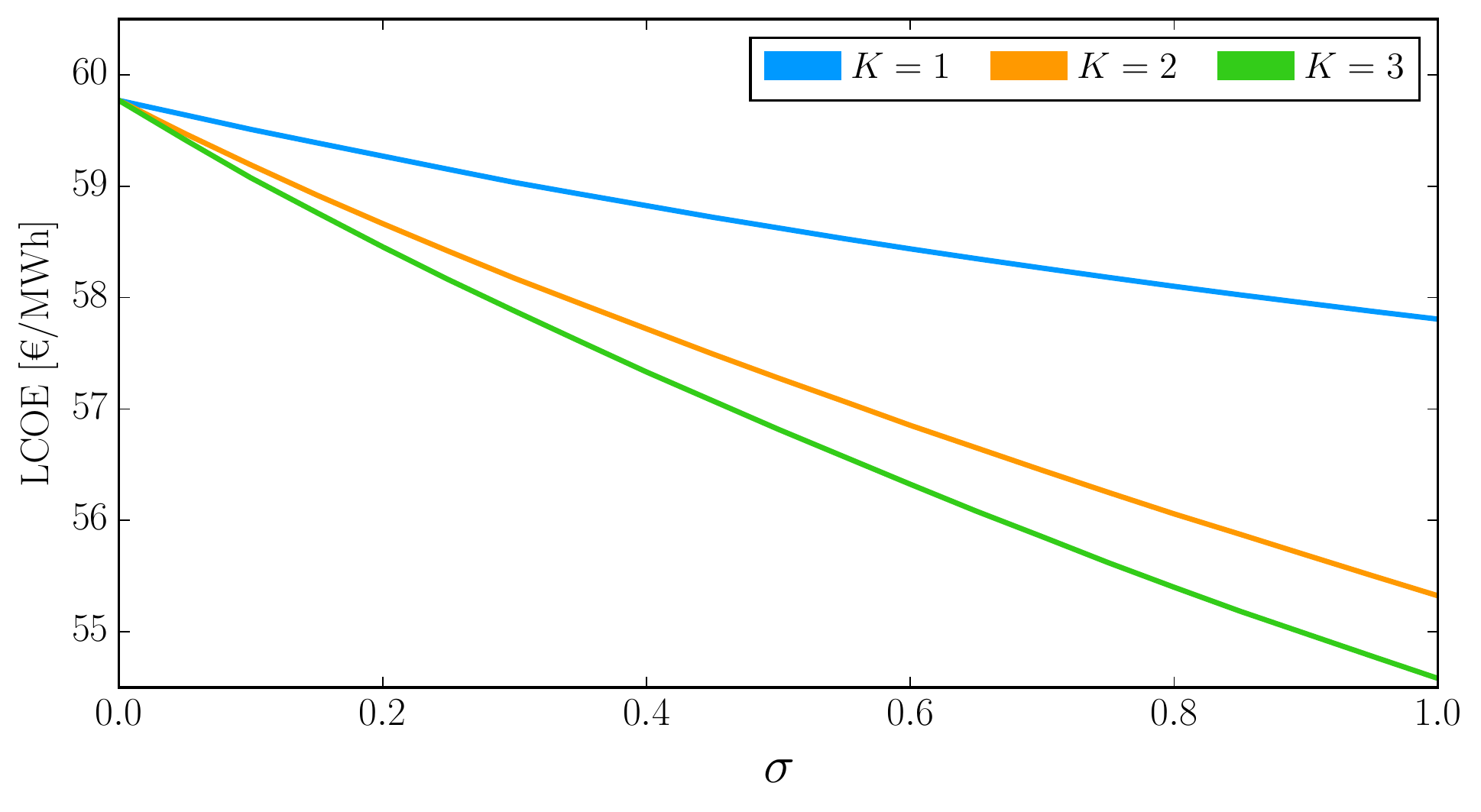}
\caption{
LCOE of the layouts interpolated between the HOM and the GAS layouts for $K = 1$ (blue), 2 (yellow) and 3 (green).
}  \label{fig:HOMinterpolationGAS}
\end{figure}

On the contrary, it might be more politically correct to reduce the heterogeneity. If the optimised GAS layouts were to represent the minimum of a rather shallow cost landscape, then other, more homogeneous layouts could be found in their vicinity without increasing the LCOE too much. Unfortunately, the search space for the exploration is high-dimensional, 60-dimensional to be more precise, as each of the 30 countries comes with its two variables $\gamma_n$ and $\alpha_n$. If for each variable we were to test two smaller and two bigger values around its GAS value, we would end up in testing $5^{60}$ layout explorations. This is infeasible. Instead, we explore simple one-parameter interpolations between the heterogenous GAS layouts and the homogeneous HOM layout:
% -------------------
\begin{eqnarray}  \label{eq:interpol}
   \gamma_n  &=&  (1-\sigma) \gamma_n^\textrm{HOM} + \sigma \gamma_n^\textrm{GAS}  \; ,  \nonumber \\
   \alpha_n     &=&  (1-\sigma) \alpha_n^\textrm{HOM} + \sigma \alpha_n^\textrm{GAS}  \; .
\end{eqnarray}
% -------------------
The interpolation parameter is confined to $0\leq\sigma\leq 1$. A value of $\sigma= 1$ represents the GAS layout while $\sigma= 0$ reproduces the homogeneous layout. Figure \ref{fig:HOMinterpolationGAS} illustrates the LCOE of the interpolated layouts. The dependence on $\sigma$ turns out to be almost linear. It is only weakly convex. This might indicate that the cost landscape around the GAS minimum is not flat, and that it might not be possible to find more homogeneous layouts without increasing the LCOE too much.

%*************************************************************************
\highlight{\section{Discussion and outlook}}
\label{sec:discussion}
%*************************************************************************

\highlight{In this paper the heterogeneity of renewable resources in different countries has been explored, but the distribution of wind and solar capacities within each country was fixed. Further heterogeneity of renewables, particularly wind, could be exploited by fine-tuning the distribution of renewables within each country, or by using a finer-scale model of Europe that exposes the locations with high capacity factors. In a recent paper \cite{Hoersch2016} it was shown that the \gls{vres} costs in a heterogenous optimisation are up to 10\% lower when using a 362 node model of Europe compared to a one-node-per-country model with 37 nodes, because the better exploitation of good sites offsets the increased exposure of grid bottlenecks within each country.}

Only three generation technologies were considered here: solar PV, onshore wind and natural gas. The inclusion of offshore wind might not improve system costs, given its high LCOE, but the LCOE may be offset by the system benefit of its steadier feed-in profile. In addition, offshore offers other benefits compared to onshore wind which are not accounted for by the cost optimisation, such as higher rates of public acceptance. \highlight{Given that offshore wind is geographically concentrated along the coastlines of countries, a finer-resolution grid model would be advisable to fully assess the integration of offshore wind.}

Modelling hydroelectricity, which already supplies 17\% of Europe's electricity, would reduce the costs of backup energy and provide extra flexibility to integrate the \gls{vres}. Similarly, the incorporation of storage or the use of flexibility from the electrification of transport and heating may alow \gls{vres} to be balanced more locally, favouring homogeneous solutions.

Finally, while the cost reduction is a strong argument for a heterogeneous \gls{vres} layout, the realisation might be a political challenge. Since the optimal placing of resources was derived from a system perspective, a realisation would require full collaboration from all countries. Countries with low capacity factors would no longer be self sufficient, while countries with high shares of renewables, such as the countries bordering the North Sea with good wind sites, may encounter problems finding enough sites or with public acceptance.

An unequal distribution of renewable energy generation also raises the question of who should pay for the generation and transmission assets. Current market conditions do not allow renewable generators to recover their capital costs from the energy-only market, forcing countries to subsidise the expansion of renewables. A highly heterogeneous system would therefore require a system for countries to compensate each other for their renewable imbalances.  Recent work on the allocation of network flows to users in highly renewable networks \cite{Brown2014,Tranberg} may provide the basis for an equitable distribution of such costs in a highly heterogeneous system.

\hspace{0.5cm}

%*************************************************************************
\highlight{\section{Conclusions}}
\label{sec:six}
%*************************************************************************

In this paper the cost-optimal spatial distribution of \gls{vres} in a simplified European electricity system has been investigated for the case where the mean \gls{vres} generation equals the mean load ($\gamma_{EU} = 1$). A heterogenous distribution of wind and solar capacities has been shown to result in an average electricity cost that is up to 11\% lower than a homogeneous distribution of renewables proportional to each country's mean load. This is because the capital costs of wind and solar dominate the total system costs, and allowing the system to build more \gls{vres} in countries with better capacity factors means that fewer wind turbines and solar panels need to be built in order to produce the same amount of energy.

If the heterogeneity parameter $K$, which controls the maximum and minimum levels of renewables generation in each country compared to its mean load, is gradually relaxed from $K=1$ (homogeneous) to larger values (heterogenous) then there is a clear trend of cost reduction, which is steepest for smaller values of $K$ and flattens out above $K=3$. This has the important policy consequence that Europe can profit from the benefits of heterogeneity without allowing renewable imbalances between countries to become excessive.

The optimal mixing parameter between wind and solar is remarkably robust as the heterogeneity is increased, favouring a high proportion of wind of between 80\% and 90\% in the \gls{vres} mix. The mixing parameter is, however, sensitive to the relative capital costs of wind and solar, dropping to between 60\% and 70\% as solar capital costs are decreased by 50\% compared to the default cost assumption.

While the best results in terms of low total system costs have been obtained here by explicit optimisation, heuristic methods for heterogeneously distributing wind and solar capacities, based for example on capacity factors, produce results that have costs only a few percent higher than the optimal systems. Given the increased comprehensibility and transparency that heuristic methods provide, this may be a price worth paying for policy makers.

\hspace{0.5cm}

\section*{Acknowledgments}

Tom Brown is funded by the CoNDyNet project, which is supported by the German Federal Ministry of Education and Research under grant no. 03SF0472C. The responsibility for the contents lies solely with the authors.

\section*{Bibliography}
\bibliographystyle{unsrt}
\bibliography{references.bib}

\end{document}